\documentclass[aps,prb,twocolumn,superscriptaddress,amsmath]{revtex4} 
\usepackage{makeidx}

\usepackage[colorlinks,
hyperindex,
colorlinks,%
citecolor=blue,%
linkcolor=blue,%
urlcolor=blue, %
]{hyperref}
\usepackage[dvipsnames]{xcolor}
\usepackage[pdftex]{graphicx}  
\usepackage{subfigure}
\usepackage{epstopdf}
\usepackage{dcolumn}   
\usepackage{bm}        
\usepackage{amssymb}   
\usepackage{amsmath}

\usepackage{siunitx}
\makeindex

\begin{document}
	
	

	\title{Spectral analysis of non-equilibrium molecular dynamics: spectral phonon temperature and phonon local non-equilibrium in thin films and across interfaces}
	
	\author{Tianli Feng}
	\affiliation{School of Mechanical Engineering and the Birck Nanotechnology Center, Purdue University, West Lafayette, Indiana 47907-2088, USA}
	
	\author{Wenjun Yao}
	\affiliation{Key Laboratory for Thermal Science and Power Engineering of Ministry of Education, Department of Engineering Mechanics, Tsinghua University, Beijing 100084, P. R. China}
	
	\author{Zuyuan Wang}
	\author{Jingjing Shi}
	\affiliation{School of Mechanical Engineering and the Birck Nanotechnology Center, Purdue University, West Lafayette, Indiana 47907-2088, USA}
	
	\author{Chuang Li}
	\affiliation{Key Laboratory for Thermal Science and Power Engineering of Ministry of Education, Department of Engineering Mechanics, Tsinghua University, Beijing 100084, P. R. China}
	
	\author{Bingyang Cao} 
	\affiliation{Key Laboratory for Thermal Science and Power Engineering of Ministry of Education, Department of Engineering Mechanics, Tsinghua University, Beijing 100084, P. R. China}
	
	\author{Xiulin Ruan}
	\email{ruan@purdue.edu}
	\affiliation{School of Mechanical Engineering and the Birck Nanotechnology Center, Purdue University, West Lafayette, Indiana 47907-2088, USA}
	\date{\today}
	

\begin{abstract}

Although extensive experimental and theoretical works have been conducted to understand the ballistic and diffusive phonon transport in nanomaterials recently, direct observation of temperature and thermal nonequilibrium of different phonon modes has not been realized. Herein, we have developed a method within the framework of molecular dynamics to calculate the temperatures of phonon in both real and phase spaces. Taking silicon thin film and graphene as examples, we directly obtained the spectral phonon temperature (SPT) and observed the local thermal nonequilibrium between the ballistic and diffusive phonons. Such nonequilibrium also generally exists across interfaces and is surprisingly large, and it provides an additional thermal interfacial resistance mechanism. Our SPT results directly show that the vertical thermal transport across the dimensionally mismatched graphene/substrate interface is through the coupling between flexural acoustic phonons of graphene and the longitudinal phonons in the substrate with mode conversion. In the dimensionally matched interfaces, e.g. graphene/graphene junction and graphene/boron nitride planar interfaces, strong coupling occurs between the acoustic phonon modes on both sides, and the coupling decreases with interfacial mixing. The SPT method together with the spectral heat flux can eliminate the size effect of the thermal conductivity prediction induced from ballistic transport. Our work shows that in thin films and across interfaces, phonons are in local thermal nonequilibrium.
\end{abstract}
\maketitle

\section{Introduction}
The transport of phonons drives the heat dissipation and shielding, thermoelectric energy conversion, energy storage and saving, etc. With the advances of nanotechnology, nowadays devices evolve towards smaller size which can be even smaller than phonon mean free path (MFP). In this case, phonons become ballistic and can travel without scattering or energy dissipation. Therefore, recently extensive experimental and theoretical work were focused on the study of ballistic and diffusive phonon transport in nanomaterials such as silicon\,\cite{Minnich2012,Regner2013nc,Hu2015nn,Zeng2015sr,Cuffe2015}, holey silicon\,\cite{Lee2015nl}, silicon nanomesh\,\cite{Zen2014nc,Maldovan2015nm}, superlattices\,\cite{Ravichandran2013nm,Maldovan2015nm,luckyanova2012coherent}, SiGe alloy nanowire\,\cite{Hsiao2013nn}, etc. By examining the size dependent thermal conductivity, researchers have found a considerable portion of ballistic phonons in the materials of several hundred nanometers to micrometer. Even though the transport properties have been extensively investigated\,\cite{Regner2013nc,Hu2015nn,Zeng2015sr,Lee2015nl,Zen2014nc,Ravichandran2013nm,Maldovan2015nm,luckyanova2012coherent,Hsiao2013nn, Feng2014Jn}, the study particularly focused on an indirect demonstration, leaving the information of phonon temperature unknown and a direct observation of the local nonequilibrium among phonons unrealized. On the other hand, the phonon transport across interface is crucially important in real devices. Even though extensive works have been conducted\,\cite{Minnich2015prb,Tian2012Enhancing, Dec2014NL,Seol2010Science,Chen2014NL,Li2012Effect,Verm2014prb,Siemens2010nm,Zhou2016Ns}, the understanding of spectral phonon interfacial conduction are still based on simple models and approximations. Thus, a direct and accurate understanding of the interfacial phonon transport is urgently needed. Recently Dunn \textit{et al.} \cite{Dunn2016} have studied the non-equilibrium of lattice vibrations in nonequilibrium molecular dynamics (NEMD). Their results are on the frequency level while not on the phonon modal level.  Another recent study\,\cite{C6CP01872F} has concluded that interfaces are at local thermal equilibrium, but the spectral nature of phonons was not considered. Here, we report a spectral phonon temperature (SPT) method that directly calculate the spatial temperatures of all the phonon modes in real systems within the framework of  NEMD, and we have directly observed the local temperature nonequilibrium among phonons in nanomaterials and across interfaces. It is demonstrated as an effective way to probe the spectral phonon thermal transport mechanisms across interface.

\section{Spectral Phonon temperature formalism}

In MD, the phonon population is described by the Boltzmann distribution
\begin{equation}
n_\lambda=\frac{k_B T_\lambda}{\hbar\omega_\lambda}.
\end{equation}
The total energy of the phonon mode $\lambda$ is the per phonon energy $\epsilon=\hbar\omega_\lambda$ multiplied by its population
\begin{equation}
\label{Etotal1}
E_\lambda=n_\lambda\epsilon= \frac{k_B T_\lambda}{\hbar\omega_\lambda}\hbar\omega_\lambda=k_B T_\lambda.
\end{equation}
Here $\lambda$ is short for $(\mathbf{k},\nu)$ with $\mathbf{k}$ and $\nu$ representing the phonon wave vector and dispersion branch, respectively. Based on the energy equipartition theorem, the time averaged kinetic energy $\langle E_{K,\lambda}\rangle$ and potential energy $\langle E_{V,\lambda}\rangle$ are both half of the total energy, e.g.,
\begin{equation}
\label{EK1}
\langle E_{K,\lambda}\rangle=\frac{1}{2}k_B T_\lambda.
\end{equation}
Based on the lattice dynamics \cite{Dove_book}, the kinetic energy of the mode $\lambda$ is
\begin{equation}
\label{EK2}
E_{K,\lambda}=\frac{1}{2}\dot{Q}_\lambda^*\dot{Q}_\lambda,
\end{equation}
where $\dot{Q}_\lambda(t)$ is the time derivative of normal mode amplitude, which is given by the Fourier transform of atomic displacement in the real space
\begin{equation}
\dot{Q}_\lambda(t) = \frac{1}{\sqrt{N_c}}\sum_{l,b}^{N_c,n} \sqrt{m_b}\exp(-i\mathbf{k}\cdot \mathbf{r}_{l,b}) \mathbf{e}_{b,\lambda}^*\cdot \dot{\mathbf{u}}_{l,b;t}.
\end{equation}
$l$ and $b$ label the indexes of the primitive cells and basis atoms with the total numbers represented by $N_c$ and $n$, respectively. $m$, $\mathbf{r}$, and $\dot{\mathbf{u}}$ are the mass, equilibrium position, and velocity vector, respectively. $\mathbf{e}_{b,\lambda}^*$ is the complex conjugate of the eigenvector component at the basis $b$ for the mode $\lambda$. 
By comparing Eq.(\ref{EK1}) and (\ref{EK2}), we can get the temperature of the phonon mode $\lambda$
\begin{equation}
\label{eq_T}
T_\lambda = \langle \dot{Q}_\lambda^*(t)\dot{Q}_\lambda(t) \rangle/k_B,
\end{equation}
where $\langle\rangle$ denotes the time average. To eliminate the fluctuation MD, Eq.(\ref{eq_T}) needs to be averaged over an enough long time.


 Note that concept of phonon normal mode is still valid under non-equilibrium \cite{Zhou2015prb1}. In Eq.(2), the temperature of a phonon mode is defined as a convenient representation of the carrier energy density, as commonly done in literature for both experimental and theoretical studies, such as Ref.\cite{Mann2006,Sullivan2017,Maassen2016,Ajit_Raman}. Another theoretical approach, spectral Boltzmann transport equation, also uses the concept of non-equilibrium temperatures of carriers such as electrons and phonons\,\cite{ni2012coupled}. Note that the temperature in MD is defined by the atomic kinetic energy ($E_K=k_BT/2$) rather than the total energy $E_K+E_V=k_BT$, although theoretically $\langle E_K\rangle=\langle E_V\rangle$.

\section{Simulation Setups}

Before applied to NEMD, our SPT method was firstly validated in equilibrium MD, in which all the phonons are in equilibrium and thus should have the same temperature. This is verified by our results in both 3D system Si and 2D system graphene (see the Appendix \ref{appendA}). 

In the following, we use our SPT method based on NEMD to study the ballistic and diffusive nature of the phonon transport in nanomaterials, e.g, Si thin film and graphene, as well as across interfaces, e.g., graphene/Si, graphene/graphene, and graphene/boron nitride. These systems cover the 2D, 3D, 2D-3D, 2D/2D junction and 2D-2D in-plane phonon transport.

\begin{figure}[t]
	\centering
	\includegraphics[width= 3.5in]{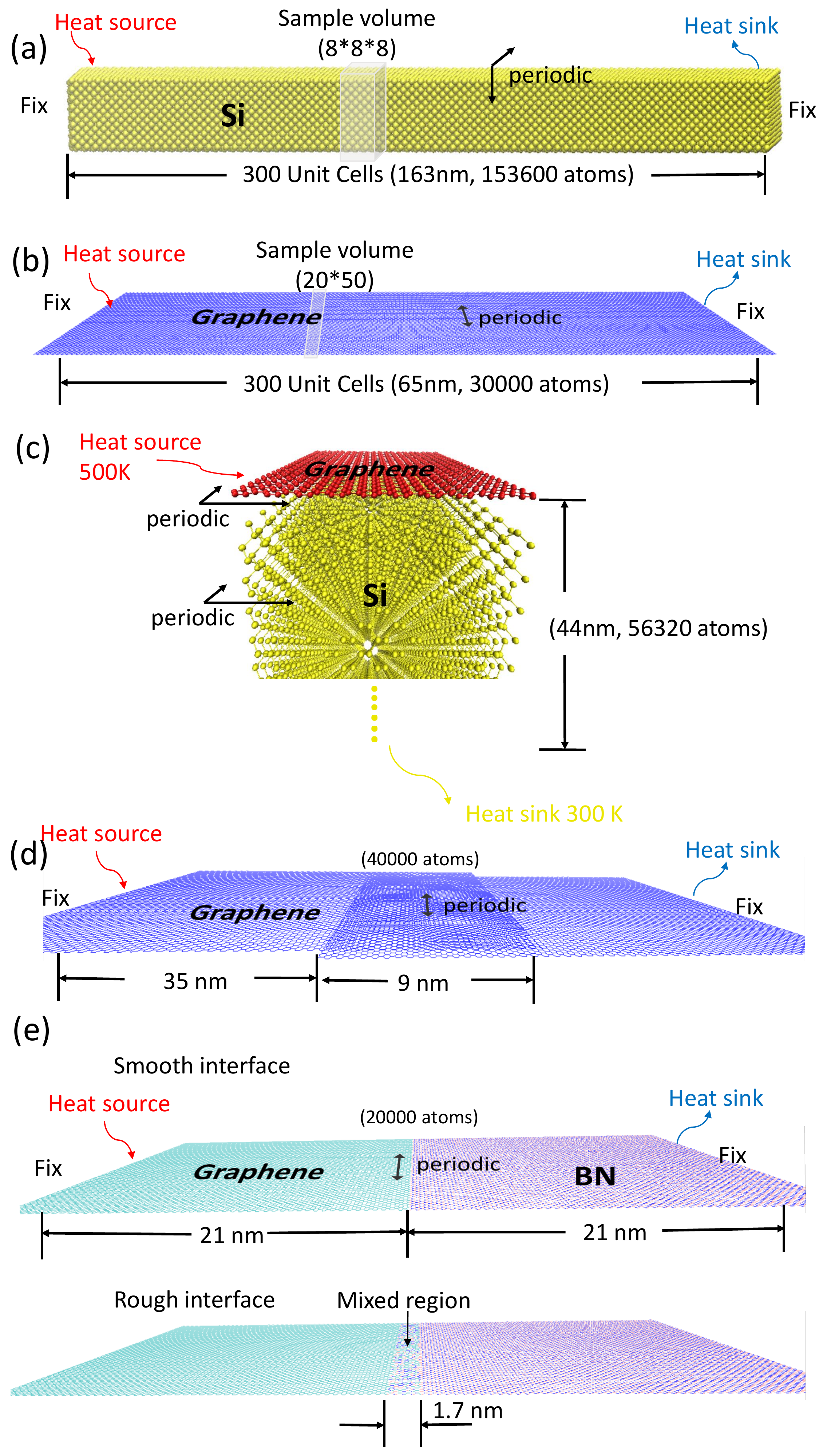}
	\caption{NEMD simulation setups for silicon (a), graphene (b), graphene/graphene vertical heat transfer (c), graphene on substrate (d), and graphene/BN (e).}\label{fig_setup}
\end{figure}

\begin{figure*}[t]
	\centering
	\includegraphics[width= 6.5in]{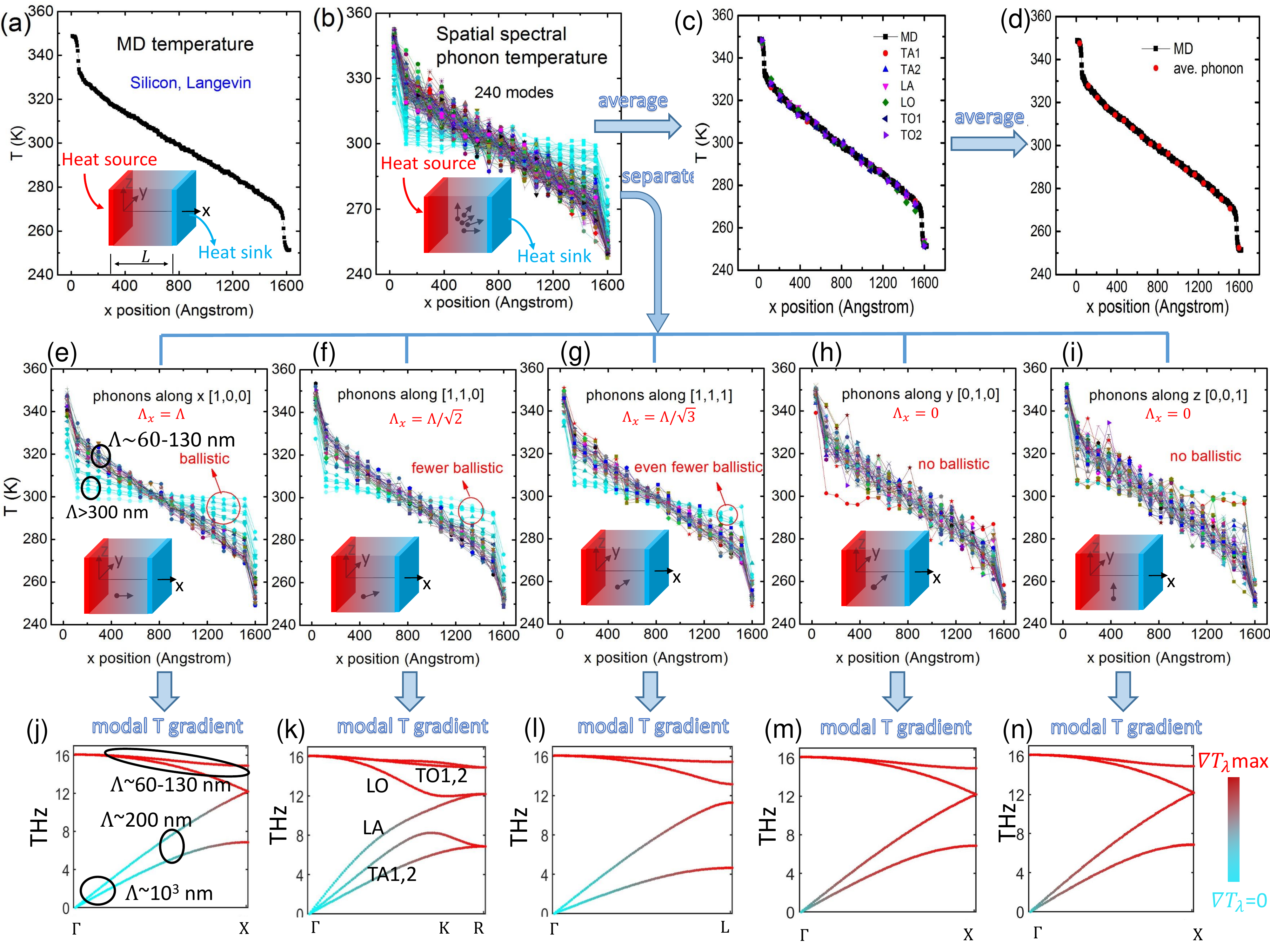}
	\caption{The spectral phonon temperatures in silicon with a temperature difference applied on the two sides in the $x$ direction. (a) The overall temperature profile in NEMD. (b) The temperatures the 240 phonon modes traveling in 5 representative directions. (c) The averaged temperatures of these phonons for the six branches. (d) The averaged phonon temperature. (e)-(f) The temperatures of the phonons traveling in the [1,0,0], [1,1,0], [1,1,1], [0,1,0], and [0,0,1] directions, respectively. (j)-(n) The phonon temperature gradient plotted into the dispersion relations with the light color representing small temperature gradient and deep color large temperature gradient.}\label{fig_Si}
\end{figure*}

The MD simulations were performed using the LAMMPS package\,\cite{LAMMPS}. The interatomic potentials used for silicon, graphene, and BN are the original\,\cite{tersoff1989modeling,tersoff1990erratum}, optimized\,\cite{Lindsay_Tersoff}, and the modified\,\cite{kinaci2012thermal} Tersoff potentials, respectively. The interactions between graphene and silicon, and between graphene layers are van der Waals (vdW) forces modeled by the Lennard-Jones (LJ) potential
\begin{equation}
\label{eq_LJ}
V(r_{ij}) = 4\epsilon \left[ \left(\frac{\sigma}{r_{ij}} \right)^{12} - \left(\frac{\sigma}{r_{ij}} \right)^{6}\right] 
\end{equation}
with the parameters $\epsilon_{C-Si}=8.909$ meV, $\sigma_{C-Si}=3.629$ \AA \cite{rappe1992uff}, $\epsilon_{C-C}=4.6$ meV and $\sigma_{C-C}=3.276$ \AA \cite{nicklow1972lattice,lindsay2011flexural}. The cutoffs for C-Si and C-C are $5\sigma$ and $3\sigma$, respectively. Here $r_{ij}$ is the distance between the atoms $i$ and $j$. The parameters can best fit the $c$-axis phonon dispersion and layer separation of 0.335 nm for graphite. The lattice constants at room temperature are obtained by relaxing the atomic structures in MD. The time step for simulations is set as 0.5 fs which is short enough to resolve all the phonon modes. For silicon, the time step can be set as 1 fs as well, and the results do not change. The two ends of the systems are fixed during the simulations as shown in Fig.\,\ref{fig_setup} (a)-(e), except for the graphene/Si system in which only the bottom boundary of silicon is fixed as shown in Fig.\,\ref{fig_setup} (d). We first relax the geometries under constant pressure and temperature (NPT) for 10 ns, and then change the ensemble to constant energy and temperature (NVE) except for the fix boundaries and heat reservoirs. After that, we apply the temperature differences on the two  reservoirs and stabilize the heat current as well as temperature gradient for 10 ns (20 million time steps). Finally the simulations are run for 20 ns to extract the atomic velocities every 10 time steps. Totally, the atomic velocities of 2 million time steps are stored and used to calculate the phonon temperature. These setups are found to be able to give stable results. The entire simulation domain is divided into many cells, and a temperature is calculated for each cell using atomic velocities in it. One such cell is the sample volume drawn in Fig.\,\ref{fig_setup}, which indicates that the atoms in that volume are used to calculate the phonon temperatures at that position. The sample volume contains $8\times8\times8$ conventional cells with 4096 atoms for silicon, and $20\times50$ with 2000 atoms for graphene and BN. To study a subtler spatial distribution of phonon temperature, a smaller sample volume can be used. Although some phonon modes are delocalized, we can still calculate their energy density inside the sample volume by transforming the real periodic vibration of the atoms to phase space. For the ballistic phonons, we indeed see that their temperature is almost a constant throughout the system, which indicates that they are indeed delocalized phonons. While for modes with short mean fee path, we can see appreciable temperature gradient. Periodic boundary conditions are applied to the lateral directions to model infinite large dimensions. The temperatures applied on the two reservoirs are 350 K and 250 K for Fig.\,\ref{fig_setup} (a) Si; 310 K and 290 K for Fig.\,\ref{fig_setup} (b) SLG; 350 K and 250 K for Fig.\,\ref{fig_setup} (c) graphene layers; 500 K and 300 K for Fig.\,\ref{fig_setup} (a) graphene on substrate; and 325 K and 275 K for Fig.\,\ref{fig_setup} (e) C/BN. Different simulation sizes, durations, temperature differences, heat reservoir time constants, and boundary conditions have been tried in our benchmark process, and the conclusions are the same. The main results shown in this paper were obtained using the simulation domain sizes and boundary conditions labeled in Fig.\,\ref{fig_setup}. The time constant for the thermal reservoirs are set as 0.5 ps, which is suitable for a stable heat current. The time constant can vary to some extent but not too much. A time constant that is too short (e.g. $<$ 0.01 ps) may lead to unstable temperature profile, while a time constant that is too long (e.g. $>$ 10 ps) is not able to build up a suitable temperature gradient.

\section{Spectral phonon temperature in silicon thin films and graphene ribbons}

\begin{figure}[!h]
	\centering
	\includegraphics[width= 3.4in]{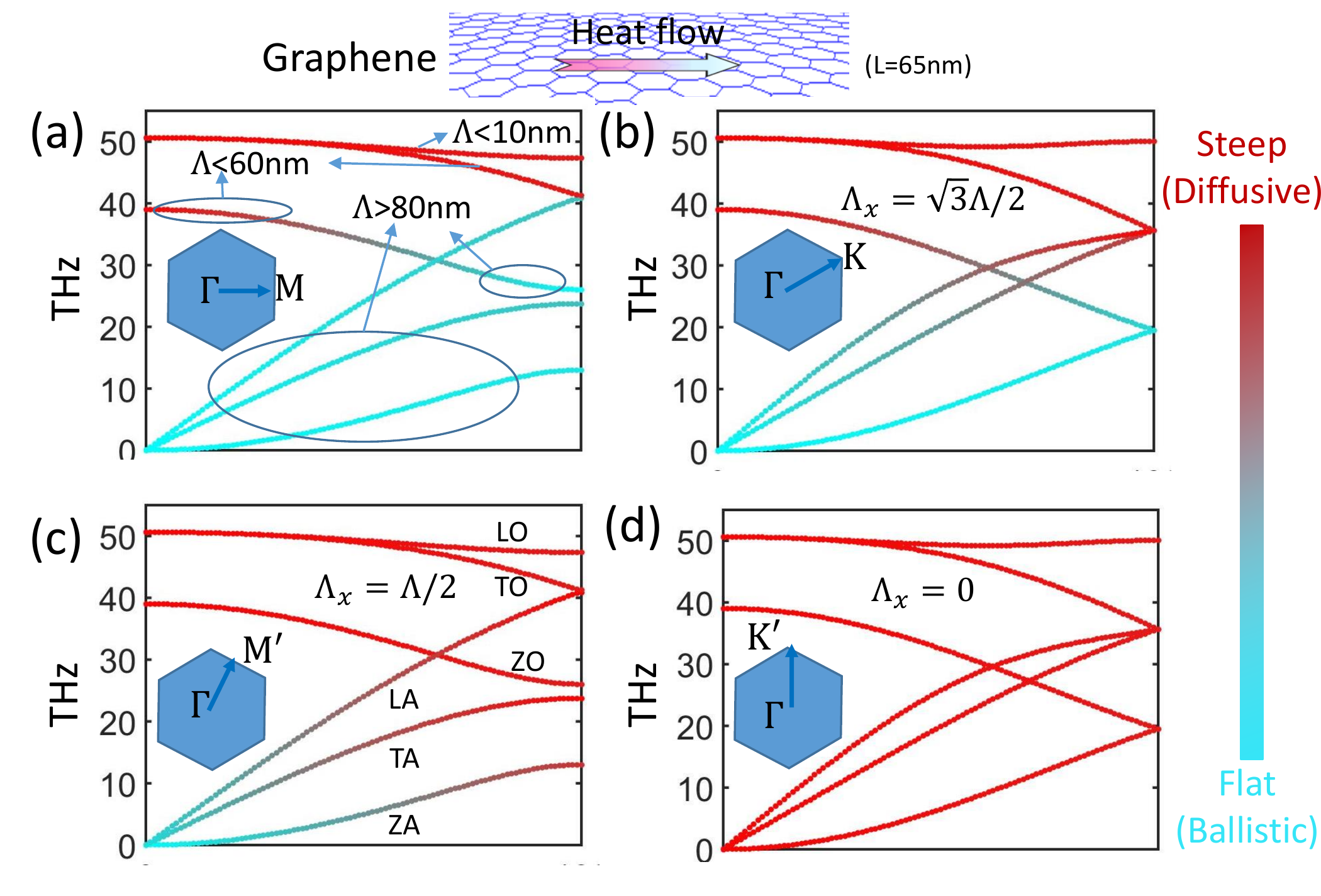}
	\caption{The spectral phonon temperature gradient of SLG with length of 65 nm under Berendsen thermostat. Light and dark colors represent small and large T gradients, respectively. Some spectral MFP values obtained by our SED method are labeled.}\label{fig_graphene}
\end{figure}

For Si thin film with the thickness $L\approx152$ nm, the left and right sides are applied with the temperatures $T_L=350$ K and $T_R=250$ K, respectively, as shown in Fig. \ref{fig_Si}. Periodic boundary conditions are applied in the $y$ and $z$ directions to mimic an infinitely large cross section. Figure \ref{fig_Si} (a) shows the overall MD temperature profile $T_{MD}$, linear in the middle and with two jumps near the thermal reservoirs, which is seen in general NEMD simulations. Then, we use the SPT method to convert $T_{MD}$ to the spectral phonon temperatures as shown in Fig. \ref{fig_Si} (b), in which we plot a total of 240 modal temperatures covering 5 representative directions, i.e., [1,0,0], [1,1,0], [1,1,1], [0,1,0] and [0,0,1]. It is seen that most phonons exhibit similar T profile as $T_{MD}$. These phonons are diffusive. Some phonon modes, however, have flatter T profiles as indicated by the light green color, and, they are ballistic phonons. To validate our method and results, we have calculated the branch-averaged temperature $\bar{T}_\nu$ and overall-averaged temperature $\bar{T}_\lambda$, using
\begin{gather}
	\bar{T}_\nu=\frac{1}{N_\mathbf{k}}\sum_\mathbf{k}^{N_\mathbf{k}} T_{\mathbf{k},\nu}=\frac{\int T_{\mathbf{k},\nu} d \mathbf{k}}{\int d \mathbf{k}}, \\
	\bar{T}_\lambda=\frac{1}{N_\lambda}\sum_\lambda^{N_\lambda} T_\lambda= \frac{1}{3n}\sum_\nu^{3n} \bar{T}_\nu,
\end{gather}
respectively (See the Appendix \ref{appendB} for the derivation). Here $n$ is the number of basis atoms, respectively. We can find that they agree well with $T_{MD}$, as shown in Figs.\,\ref{fig_Si} (c) and (d). To gain more insight, the temperatures of the phonons traveling in the 5 representative directions are separately shown in Figs.\,\ref{fig_Si} (e)-(i). For the phonons traveling along the x direction, nearly half of the modes are (quasi)ballistic. We calculate the MFP of these phonons using a separate independent method -- spectral energy density (SED) analysis based on EMD (Appendix \ref{appendC}), and find that their MFP are generally longer than the Si thin film thickness 152 nm. The other modes, which are diffusive, are found to have considerably shorter MFP, 60-130 nm, as indicated in Fig.\,\ref{fig_Si} (e). Such an agreement between our NEMD-based SPT results and the EMD-based SED results strengthens the validity of the SPT method. These results agree qualitatively well with recent experimentally reconstructed MFP spectra of silicon\,\cite{Minnich2012,Cuffe2015,Hu2015nn,Zeng2015sr,Regner2013nc}, while providing more direct observation of ballistic and diffusive phonons whose temperatures have not been measured yet. For the phonons traveling in other directions, their effective phonon MFP is that projected in the x direction $\Lambda_x$. Therefore, oblique the traveling directions are, the more diffusive the phonon modes become. For example, in the [1,1,0] direction, $\Lambda_x=\Lambda/\sqrt{2}$, and more diffusive modes appear. In [0,1,0] and [0,0,1] directions, $\Lambda_x=0$ and all modes are diffusive. Note that in Figs.\,\ref{fig_Si} (e)-(i), a few long-wavelength modes have large uncertainty due to the limitation of MD. To see clearly which phonon modes are ballistic, we designate the modal temperature gradient in the phonon dispersion relation as shown in Figs.\,\ref{fig_Si} (j)-(n). It is seen that the fraction of ballistic modes decreases as phonon traveling direction becomes more oblique. We also tried the Berendsen thermostat and got similar results as shown in Appendix \ref{appendD}.

The SPT method also provides a way to calculate the phonon MFP spectra. According to the phonon Boltzmann transport equation (BTE), the phonons' temperature jumps $\Delta T_\lambda$ near the contacts are related to their MFP $\Lambda$ by\,\cite{Jeong2010jap, Lundstrom2016} 
$\frac{T_L-T_R}{2\Delta T_\lambda}=1+L/(\frac{4}{3}\Lambda_\lambda)$.

We have also obtained the modal temperatures of a 2D system, single-layer graphene, with length of 65 nm. As shown in Fig.\,\ref{fig_graphene}, along $\Gamma$-M the acoustic phonons are ballistic while most optical phonons are diffusive. As the direction deviates from the heat flow direction, the portion of ballistic phonons reduces. Along $\Gamma$-K' which is perpendicular to the heat flow, nearly no phonon is ballistic. Since most conclusions are similar to those of Si thin film, we will not reiterate them in the main text (See Appendix \ref{appendE} for details). 

We note that in both Si thin films and graphene ribbons, the assigned temperature can be well maintained deep inside the reservoir, while cannot be maintained in the region of the reservoir near the channel. As a result, the modal temperature is in equilibrium deep inside the reservoirs while in non-equilibrium near the channel. Dunn \textit{et al.} \cite{Dunn2016} found phonon non-equilibrium in the reservoirs as they sampled the whole reservoirs to obtain the phonon information. This result is consistent with ours.



\section{Spectral Phonon temperature and phonon local non-equilibrium across Si/graphene interface}

\begin{figure}[t]
	\centering
	\includegraphics[width= 3.3in, height=5.5in]{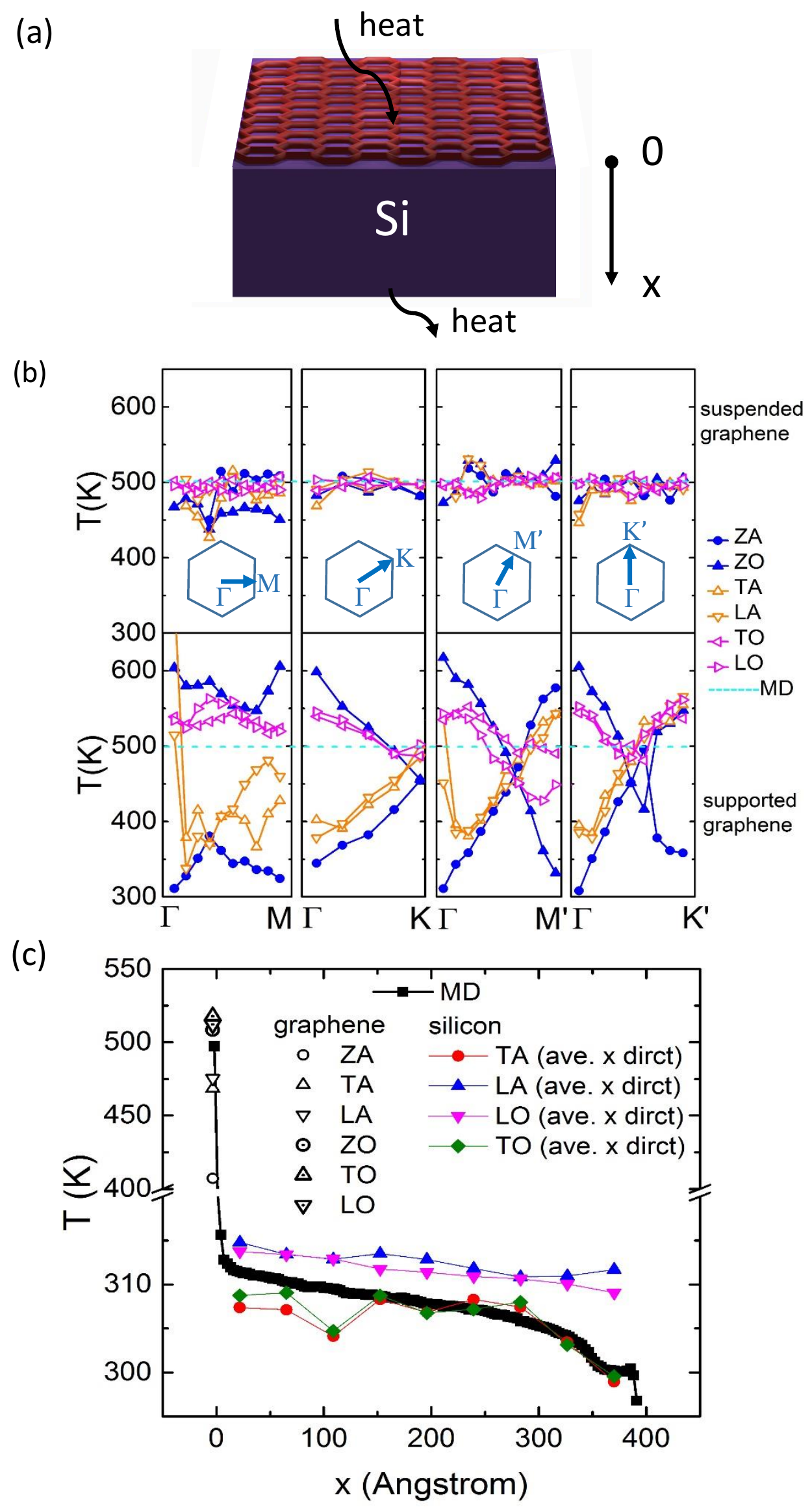}
	\caption{(a) A schematic of the graphene/Si vertical thermal transport, with detailed information shown in Fig.\cal\ref{fig_setup} (c). (b) The spectral temperatures of the phonons traveling in 4 directions in suspended graphene and supported graphene. (c) The branch temperatures of graphene and the silicon substrate with comparison to $T_{MD}$.}\label{fig_GSi}
\end{figure}

We now investigate the thermal transport across interfaces. Due to the 2D nature, graphene is typically supported on a substrate  in real devices. The presence of the substrate can affect the electronic structure, phonon dispersion\,\cite{Zhou2007Nm,Ong2011prb,Qiu2012apl2}, and transport properties\,\cite{Fratini2008prb,Seol2010Science,Chen2008Nn,Morozov2008prl,Qiu2012apl2}. For example, the mobility can be reduced by a factor of 10\,\cite{Fratini2008prb} and the thermal conductivity reduced by a factor of 5\,\cite{Seol2010Science}. The heat dissipation from graphene to substrate is an important thermal management issue in devices\,\cite{Freitag2009NL,Yan2012NC}. Although it is well known that the flexural modes couple most strongly with the substrate\,\cite{Qiu2012apl2,Seol2010Science}, it has been an open question, however, on which phonon modes in substrate they couple to\,\cite{Sadeghi2013NL}. We tackle this problem using our spectral phonon temperature method.
Figure \ref{fig_GSi} (a) shows the schematic of the heat conduction from graphene at 500 K to the Si substrate with bottom temperature maintained at 300 K. Periodic boundary conditions are applied to the in-plane directions to eliminate nano-size effect. As a control case, a suspended graphene with T = 500 K is also studied. The wavevector-resolved  spectral phonon temperatures in suspended and supported graphene are shown in Fig. \,\ref{fig_GSi} (b) panels. Clearly the phonons are in thermal equilibrium in suspended graphene as expected, surprisingly they are in strong nonequilibrium in supported graphene. The branch temperatures of supported graphene and the substrate are shown in Fig. \,\ref{fig_GSi} (c) with comparison to $T_{MD}$. 
In graphene, the out-of-plane acoustic (ZA), transverse acoustic (TA) and longitudinal acoustic (LA) phonons have much lower temperatures than the transverse optical (TO), longitudinal optical (LO) and out-of-plane optical (ZO) phonons. ZA is the lowest among them. 
While on the surface of the Si substrate, the LA and LO branches have much higher temperature than the other modes. This indicates the most efficient interfacial transport channel is that the acoustic branches especially the ZA phonons in graphene transmit into the LA and LO branches in the substrate.
The ZA mode in graphene couples strongly with the substrate because its out-of-plane atomic vibrations directly "press" the substrate and excite the substrate LA and LO modes, which have the same out-of-plane atomic motion. Phonon mode conversion behavior is surprising and cannot be obtained using other methods such as acoustic or diffuse mismatch methods. 
Interestingly, the ZO mode, which was believed to couple strongly with the substrate for the in-plane thermal transport\cite{Qiu2012apl2,Seol2010Science}, does not contribute much to the vertical thermal transport. This is probably due to that its frequency is higher than the phonon cutoff frequency in silicon, and the elastic transmission process does not allow its transmission.
Not only different branches behave differently, but inside a branch the phonon with different wavevectors are in quite large nonequilibrium, as shown in Fig.\,\ref{fig_GSi} (b). The lower-energy acoustic phonons typically have much lower temperature, and thus couple more strongly with the substrate, which can be due to their higher transmission to the substrate and weaker coupling to the other hot phonon modes in graphene\,\cite{Ajit2016prb}. 

\begin{figure}[t]
	\centering
	\includegraphics[width= 3.4in]{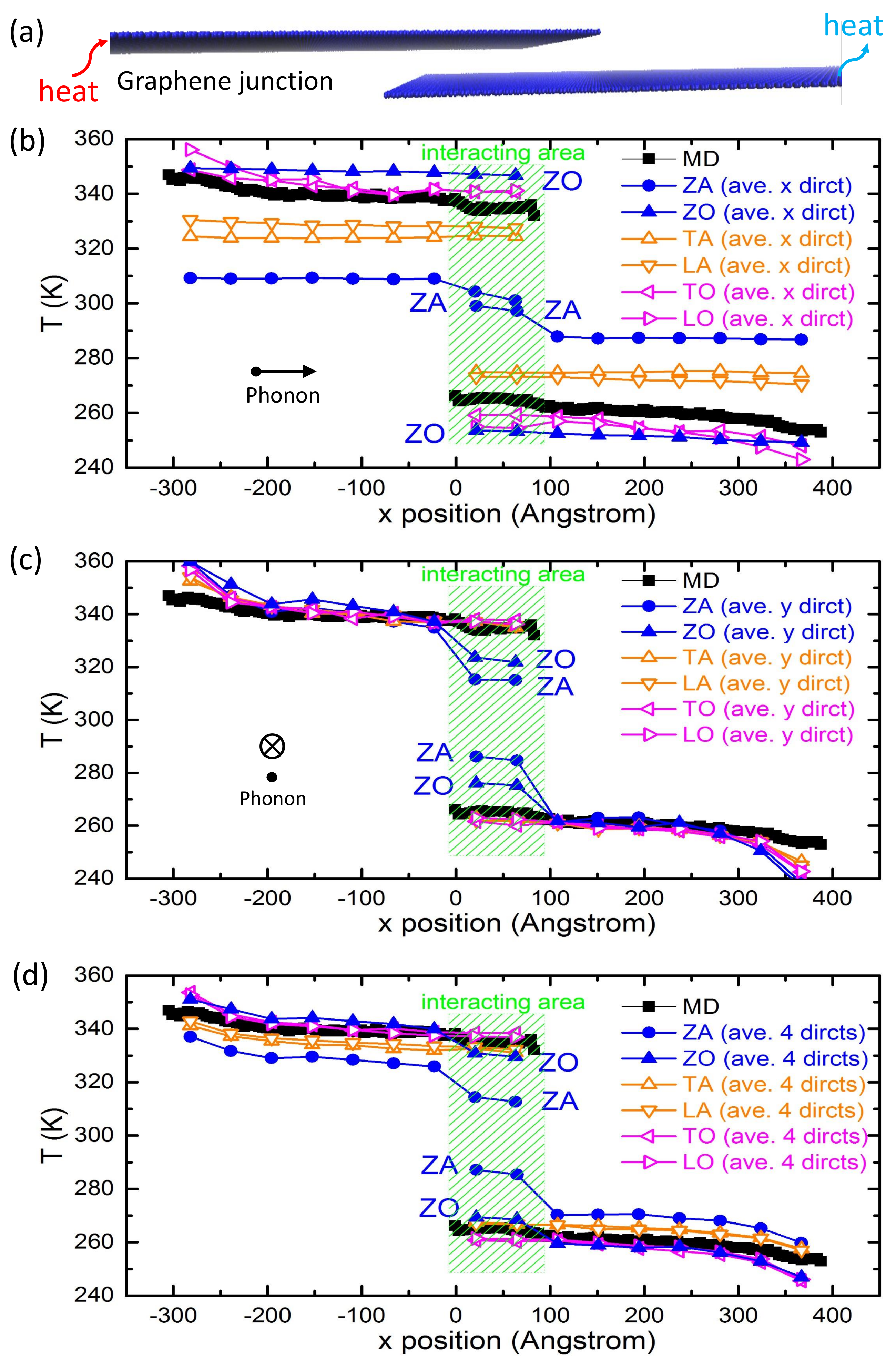}
	\caption{A sketch of the graphene/graphene junction heat flow (a). $T_{MD}$ with comparison to the spatial branch temperature of the phonons traveling in the $x$ (b), $y$ (c), and averaged 4 directions (d).}\label{fig_Glayer}
\end{figure}

\begin{figure*}[t]
	\centering
	\includegraphics[width= 6.1in]{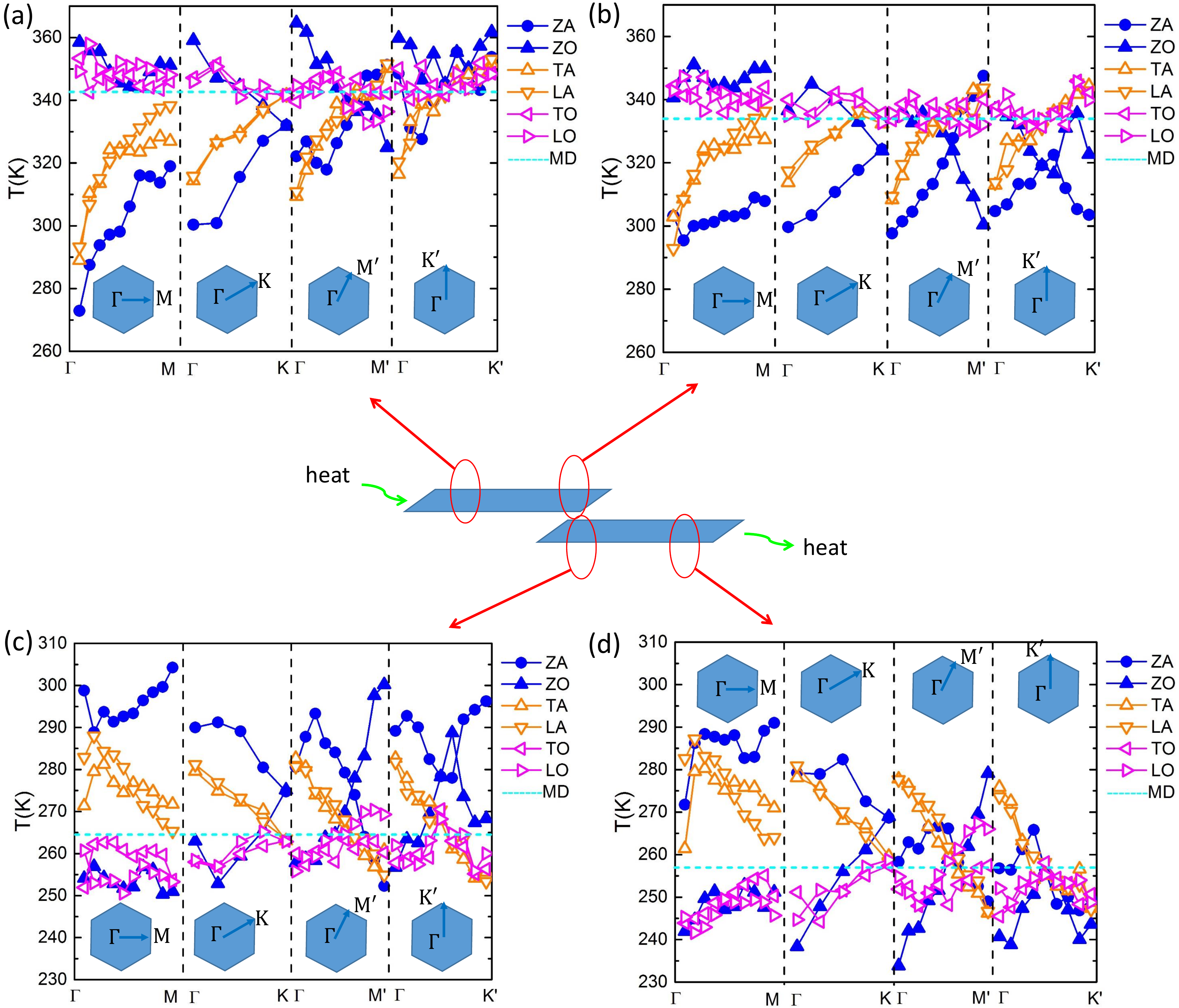}
	\caption{The spectral phonon temperatures $T_{\lambda, x}$ at four representatively positions: (a) the upper layer far from the G/G junction, (b) the upper layer part of the G/G junction, (c) the lower layer part of the G/G junction, (d) the lower layer far from the G/G junction. At each position, the phonons with wavevectors in four directions, $\Gamma-M$, $\Gamma-K$, $\Gamma-M'$, and $\Gamma-K'$ are studied. The heat flow is parallel to the $\Gamma-M$ direction. As a reference, the green dash line shows the temperature from MD at each position. We can clearly observe the temperature change with branches, traveling directions, wavelength, and positions. It is seen that the longer-wavelength acoustic phonons (ZA, TA and LA) have lower temperature in the upper layer and higher temperature in the lower layer.}\label{fig_glayer_T}
\end{figure*}

We find that the substrate also breaks the in-plane symmetry of graphene, in addition to the breakdown of its out-of-plane symmetry that was discovered previously\,\cite{Qiu2012apl2}. For example, the spectral phonon temperatures of graphene along $\Gamma$-$M$ become quite different from those along $\Gamma$-$M'$ when a substrate is applied as shown in Fig.\,\ref{fig_GSi} (b). This is because graphene has a sixfold axis of rotation while Si's is fourfold. The atoms' vibrations along $\Gamma-M$ in graphene align with those along [0,0,1] in silicon, while those along $\Gamma-M'$ in graphene aligns with [0,2,$\sqrt{3}$] in silicon. This finding may be used to engineer the thermal transport between graphene and substrate using alignment.


Moreover, the phonon thermal nonequilibrium near interface creates a new resistance mechanism. Heat needs to be transferred from the optical modes to the acoustic modes in graphene first; then to the longitudinal modes in the substrate; and finally back to other modes in the substrate. This is in analogy to electron-phonon coupled thermal transport across metal-dielectric interfaces, where electron-phonon nonequilibrium in metals introduces an additional resistance \cite{Majumdar2004apl,Wang2016jap}.

\section{Phonon local non-equilibrium in graphene-graphene 2D junction}

Here, we investigate a 2D-2D junction, the graphene-graphene junction, which is common in graphene foams that are promising for thermal management and energy storage\,\cite{Zhao2014EES,Zhou2013AFM,Han2014AM,Huang2014AM,Thiyagarajan2014APL}. As indicated in Fig.\,\ref{fig_Glayer} (a), the heat flows from the left side of the top graphene to the bottom graphene with the cross-plane distance 0.335 nm (AB stack), same as in graphite. Inside the graphene layers away from the junction, the MD temperature profile is quite smooth and flat due to the strong sp${}^2$ bond and ballistic transport nature. Near the junction, at $x=0$ for the top layer and $x=100$ \AA \ for the bottom layer, there are small jumps, indicating that the interlayer van der Waals field indeed drives the heat flow.

To gain insight of the spectral phonon heat conduction channel, we calculated the modal temperatures as shown in Fig.\,\ref{fig_Glayer} (b)-(d). First, we inspect the phonons traveling along the $x$ (heat flow) direction Fig.\,\ref{fig_Glayer} (b). Surprisingly, the phonon local thermal nonequilibrium is quite large throughout the whole system. At the junction the spectral phonon non-equilibrium is due to the different spectral heat transfer efficiency, while away from the junction the spectral non-equilibrium is due to the ballistic effect since the length is much shorter than phonon MFP in graphene. Clearly, the three acoustic phonon branches have much smaller jumps than the optical phonon branches, and they dominate the heat transfer. Among them, ZA mode is the most efficient one, whose the temperature jump is about only 10\% of the overall $\Delta T_{MD}$. The out-of-plane vibration of the ZA mode can directly transfer energy across the vertical junction. The optical branches, including the ZO mode, however, cannot transfer heat efficiently, probably due to the negative group velocity with respect to the heat flow direction. 

Now we inspect the phonons traveling in the $y$ direction, perpendicular to the heat flow, as shown in Fig.\,\ref{fig_Glayer} (c). These phonons do not contribute to the thermal transport which is along the $x$ direction. We find a completely different phenomenon compared to that in the thermal transport direction. Here the phonon temperatures match well with the MD temperature except for the junction region. This is because that these phonons are diffusive as we discussed previously, and the phonon non-equilibrium away from the junction does not exist anymore. While at the junction, the spectral non-equilibrium still exist due to the different spectral heat transfer efficiency. Among these phonons, ZA and ZO branches have the smallest temperature jumps, probably due to their out-of-plane vibration's governing the interlayer interaction. The branch temperatures averaged among 4 representative directions are plotted in Fig.\,\ref{fig_Glayer} (d). 

Furthermore, based on the $\mathbf{k}$-resolved phonon temperature shown in Fig.\,\ref{fig_glayer_T}, we find that the longer-wavelength acoustic phonons have lower temperature in the upper layer and higher temperature in the lower layer, while the optical phonons are opposite. This indicates that the longer-wavelength acoustic phonons contribute more to the junction heat transfer.

\section{Phonon local non-equilibrium in graphene-boron nitride 2D interface}

\begin{figure}[tbph]
	\centering
	\includegraphics[width= 3.4in]{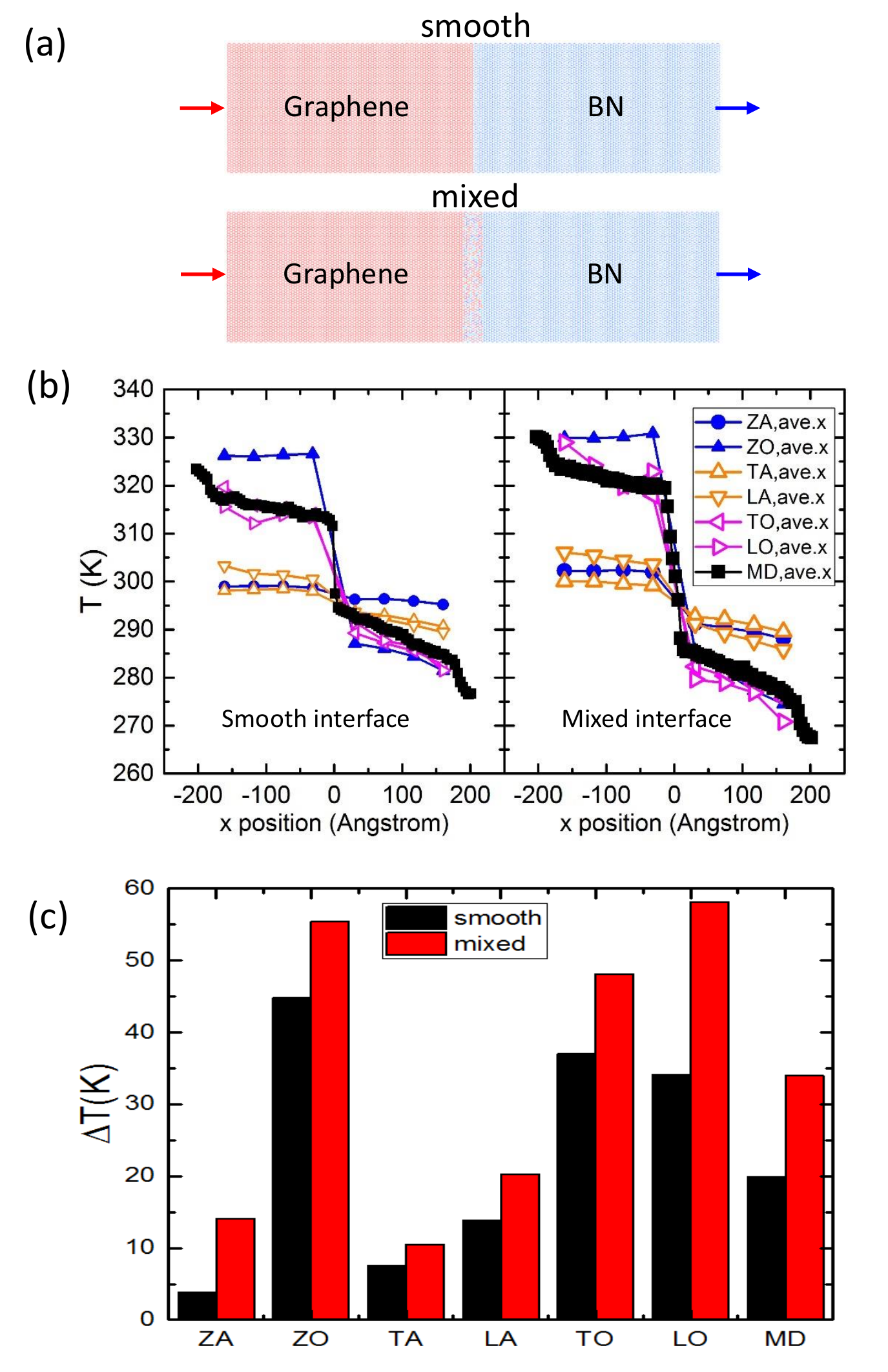}
	\caption{(a) A sketch of graphene/BN interfaces, with detailed information shown in Fig.\,\ref{fig_setup} (e). (b) The $T_{MD}$ with comparison to the spatial branch temperature of the phonons traveling in the $x$ direction. (c) The phonon branch temperature jumps.}\label{fig_CBN}
\end{figure}

\begin{figure}[tpbh]
	\centering
	\includegraphics[width= 3.5in]{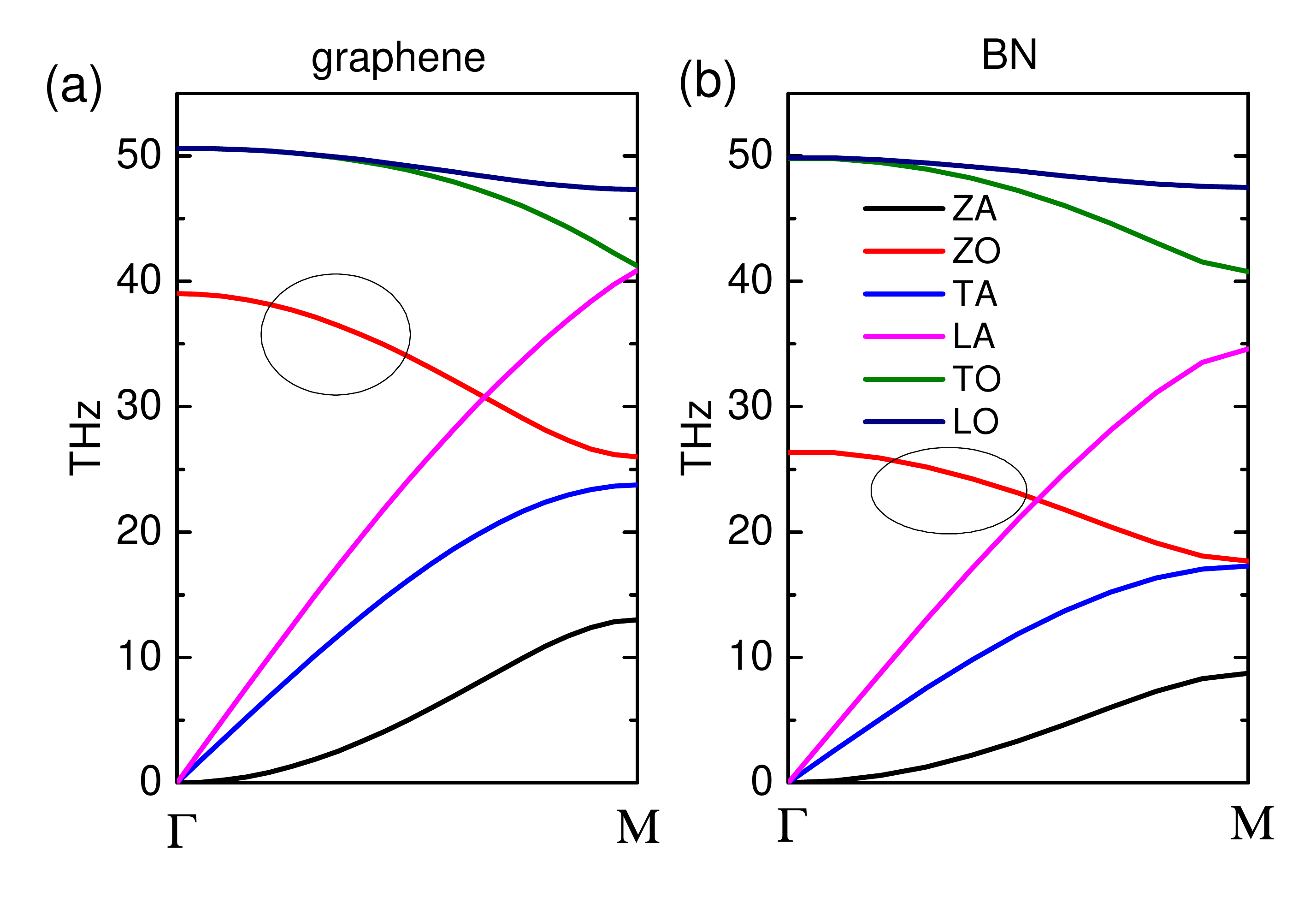}
	\caption{The phonon dispersion relations of graphene (a) and hexagonal boron nitride (b) from $\Gamma$ to $M$. The ZO branches has negligible frequency overlap.}\label{fig_disp_BN}
\end{figure}

After inspecting the dimensionally mismatched interfaces and matched junctions, now we take the graphene/boron nitride interface Fig.\,\ref{fig_CBN} (a) as an example to examine the 2D planner interfacial thermal transport. Surprisingly, even across such a nearly perfect-lattice-matched interface, the phonon local thermal nonequilibrium is quite large as shown in Figs.\,\ref{fig_CBN} (b) and (c). The coupling between the acoustic phonon modes ZA, TA and LA is quite strong, and the temperature jump of the ZA mode is only 10-20\% of the overall $\Delta T_{MD}$. One interesting phenomenon is that the temperature jump of the ZO mode is as high as 220\% of the $\Delta T_{MD}$ although it vibrates in the out-of-plane direction as the ZA mode. Such large jump and thus inefficient heat conduction are due to the negligible phonon band overlap between graphene and BN as shown in Fig.\,\ref{fig_disp_BN}, which leads to negligible elastic transmission. Furthermore, if interface mixing is introduced, $\Delta T_{MD}$ as well as all the phonon modal temperature jumps $\Delta T_{\lambda}$ increase. Interestingly, relatively the jump of the ZA mode increases the most, from 19\% to 41\%, due to the break of the out-of-plane symmetry, while the relative jumps of the in-plane modes decreases, demonstrating that the heat transfer channel shifts a certain amount from ZA mode to the TA and LA modes. 



%
%

\section{Spectral phonon thermal conductivity}

\begin{figure*}[t]
	\centering
	\includegraphics[width= 6.5in]{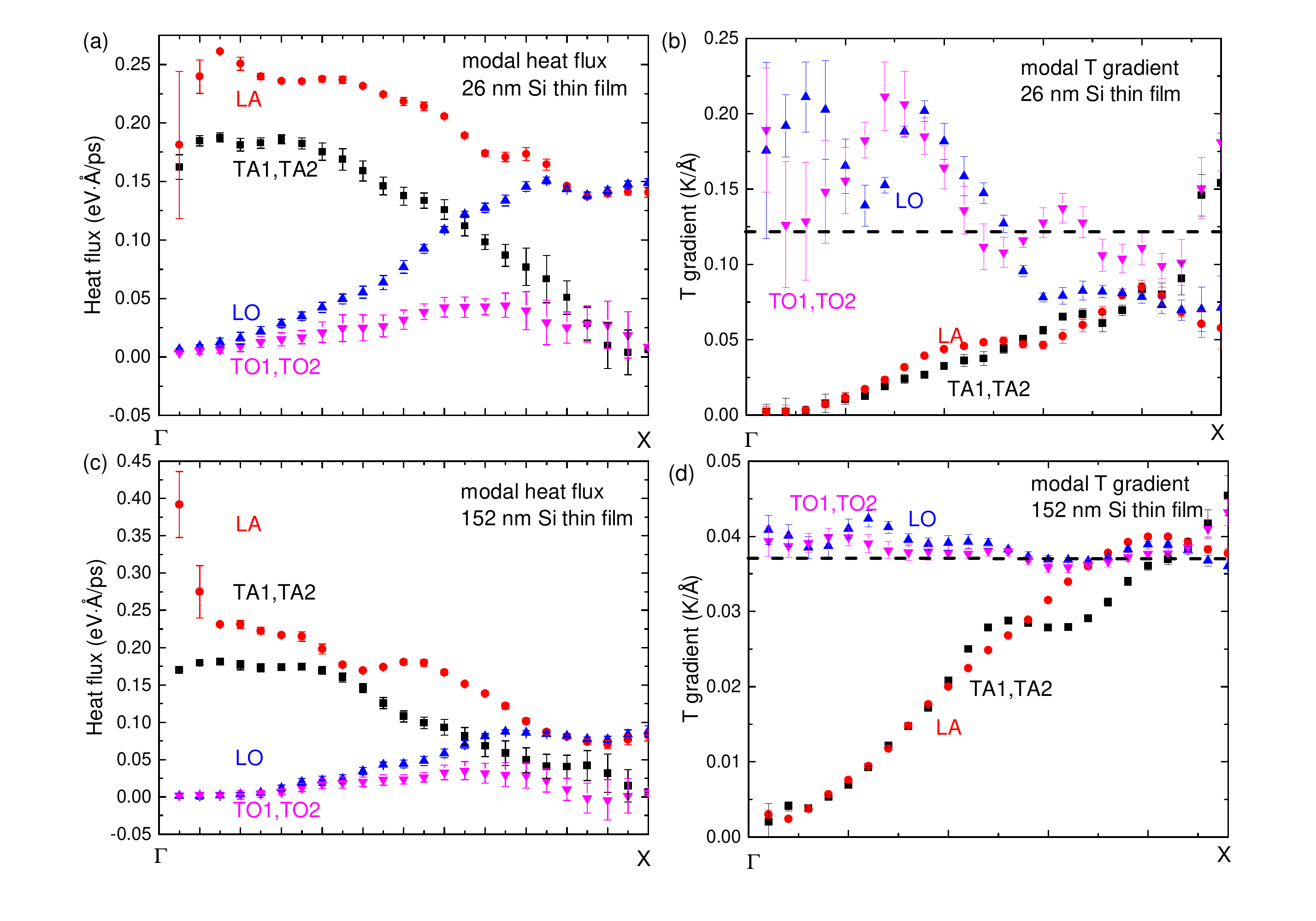}
	\caption{The spectral phonon heat flux and temperature gradient from $\Gamma$ to X in 26-nm (a,b) and 152-nm (c,d) Si thin films. The dash lines indicate the gradient of $T_{\rm MD}$.}\label{fig_modalHF}
\end{figure*}

With our SPT method, the spectral phonon thermal conductivity $\kappa_\lambda$ that is defined by Fourier's law
\begin{equation}
\kappa_\lambda=\frac{q_\lambda}{\nabla T_\lambda},
\end{equation}
can be obtained, where the spectral heat flux $q_\lambda$ can be calculated by the method developed by Zhou and Hu \cite{Zhou2015prb1} recently.
\begin{equation}
q_\lambda=\sum_{l,b}^{N_c,n}\langle \frac{1}{\sqrt{N_cm_b}}\left[E_{l,b}(t)-\mathbf{S}_{l,b}(t)\right]\mathbf{e}_{b,\lambda}\exp(i\mathbf{k}\cdot\mathbf{r}_{l,b})\dot{Q}_\lambda\rangle.
\end{equation}
Here $E_{l,b}$ and $\mathbf{S}_{l,b}$ are the total energy and stress of the atom (l,b), respectively. We take 26-nm and 152-nm  Si thin films as examples, and calculate the spectral phonon heat flux and temperature gradient along $\Gamma$-X as shown in Fig.\,\ref{fig_modalHF}. It is clearly seen that the TA and LA branches carry the most amount of heat (Fig.\,\ref{fig_modalHF} a, c) although their temperature gradients are the smallest (Fig.\,\ref{fig_modalHF} b, d). This certainly indicates that their spectral thermal conductivities are the largest. Generally, the heat fluxes of all the acoustic and optical branches decrease with increasing phonon frequency, which agree well with decreasing trend of phonon relaxation time with frequency. At X point, the modal heat flux $q_{\rm LA}$ merges with $q_{\rm LO}$, and $q_{\rm TA}$ merges with $q_{\rm TO}$. The former is due to the degeneration of the LA and LO branches at X point, and the latter is due to the zero group velocity of both TA and TO branches at X point. Overall, the error bars of the results are small, considering that molecular dynamics is statistical. Relatively, the uncertainty of optical phonons are larger than that of acoustic phonons due to the diffusive nature. But this is tolerable since optical phonons only contribute a small portion to the total thermal conductivity.

\begin{figure}[tpbh]
	\centering
	\includegraphics[width= 3.5in]{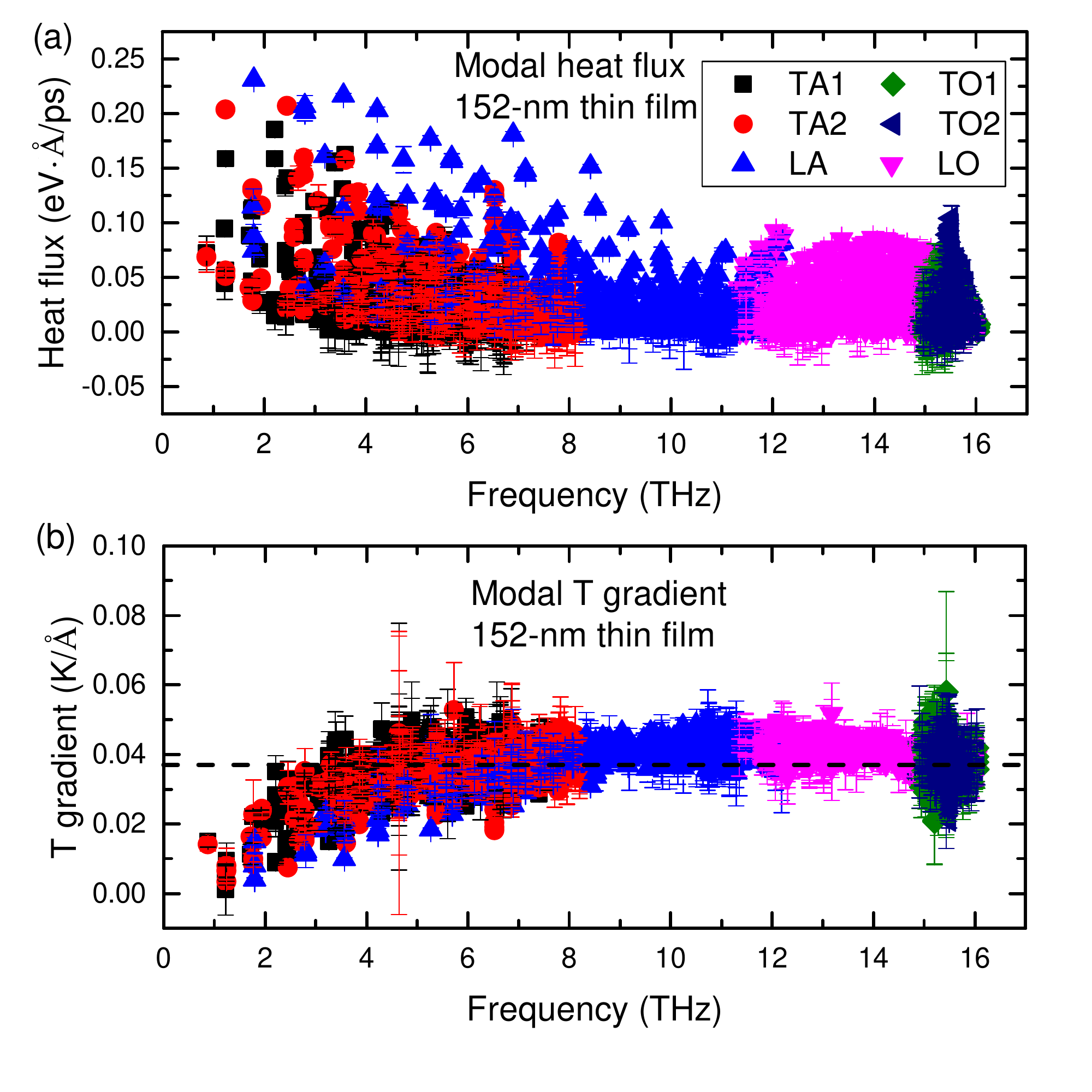}
	\caption{The spectral phonon heat flux and temperature gradient of the first octant of the Brillouin Zone with $16\times16\times16$ k-mesh in the 152-nm Si thin film. The dash line indicates the gradient of $T_{\rm MD}$.}\label{fig_modalBZ}
\end{figure}

To further verify the accuracy of our results, we calculated the modal heat flux and temperature gradient in the whole octant of the Brillouin Zone with $16\times16\times16$ k-mesh as shown in Fig.\,\ref{fig_modalBZ}. We have found that the total heat flux $\sum_{\lambda}q_\lambda$ and the average temperature gradient $\nabla\bar{T}_\lambda$ agree excellently with the MD heat flux $q_{\rm MD}$ and temperature gradient $\nabla T_{\rm MD}$, respectively.

The conventional thermal conductivity calculated from the grey temperature $T_{\rm MD}$
\begin{equation}
\label{eq_kappa}
\kappa=\frac{q_{\rm MD}}{\nabla T_{\rm MD}}=\frac{\sum_{\lambda}q_\lambda}{\nabla T_{\rm MD}},
\end{equation}
misses the spectral phonon temperature information, and thus underestimates the thermal conductivity contribution of the ballistic modes. For the 152-nm Si thin film, Eq. (\ref{eq_kappa}) yields 148 W/mK based on the modal heat flux$\sum_{\lambda}q_\lambda$ which agree well with the 151 W/mK calculated from direct MD heat flux $q_{\rm MD}$. After considering the spectral phonon temperature, the thermal conductivity 
\begin{equation}
\kappa=\sum_\lambda \frac{q_\lambda}{\nabla T_\lambda},
\end{equation}
is obtained as 178 W/mK which eliminates the size effect induced from ballistic transport. We note that it is still lower than the bulk value 235 W/mK determined by the NEMD length extrapolation based on Tersoff potential. This is probably due to the reflection at the boundaries near the reservoirs.

\section{Conclusions}
In summary, we have developed a powerful SPT method to extract the spectral phonon temperature directly from atomistic simulations, and demonstrated its capability of resolving ballistic and diffusive phonon modes in nanomaterials and spectral phonon mode coupling across interfaces. In the benchmark materials, nano-size Si and graphene, the thermal local nonequilibrium between the ballistic and diffusive phonons has been clearly observed. Such nonequilibrium exists in interfaces as well and is surprisingly large, based on our study of graphene/substrate, graphene/graphene junction and graphene/BN planar interface. The phonon local thermal nonequilibrium introduces a new mechanism of thermal interfacial resistance. In particular, the most efficient thermal transport channel across the dimensionally-mismatch graphene-substrate interface is the transmission of in-plane acoustic modes in graphene into the cross-plane LA and LO modes in the substrate. The interface roughness is shown to substantially affect such nonequilibrium and coupling, and shift the heat transfer channel. The SPT method together with the spectral heat flux can predict the thermal conductivity with the size effect induced from ballistic transport eliminated. Since our method can directly extract the spectral and spatial phonon temperature from MD simulations, it is expected to have broad applicability to many thermal applications such as thermal diodes, thermal interface materials (TIMs), coherent phononic meta-materials, etc.

\begin{acknowledgements}
Simulations were preformed at the Rosen Center for Advanced Computing (RCAC) of Purdue University. The work was partially supported by the National Science Foundation (Award No. 1150948) and the Defense Advanced Research Projects Agency (Award No. HR0011-15-2-0037). We would like to thank Yanguang Zhou for helpful discussions on the spectral heat flux decomposition.
\end{acknowledgements}



\appendix
\renewcommand\thefigure{\thesection.\arabic{figure}}

\section{Benchmark the method in equilibrium molecular dynamics}
\setcounter{figure}{0}
\label{appendA}
This section is to (1) determine the total simulation time to obtain good modal temperature results, (2) determine a better approach between the long-time simulation and the ensemble average of short-time simulations since the latter is more time-efficient, and (3) verify that our spectral phonon temperature method does not require periodic boundary conditions that is required for the normal mode analysis to calculate phonon relaxation time.

\begin{figure}[!h]
	\centering
	\includegraphics[width= 3.5in]{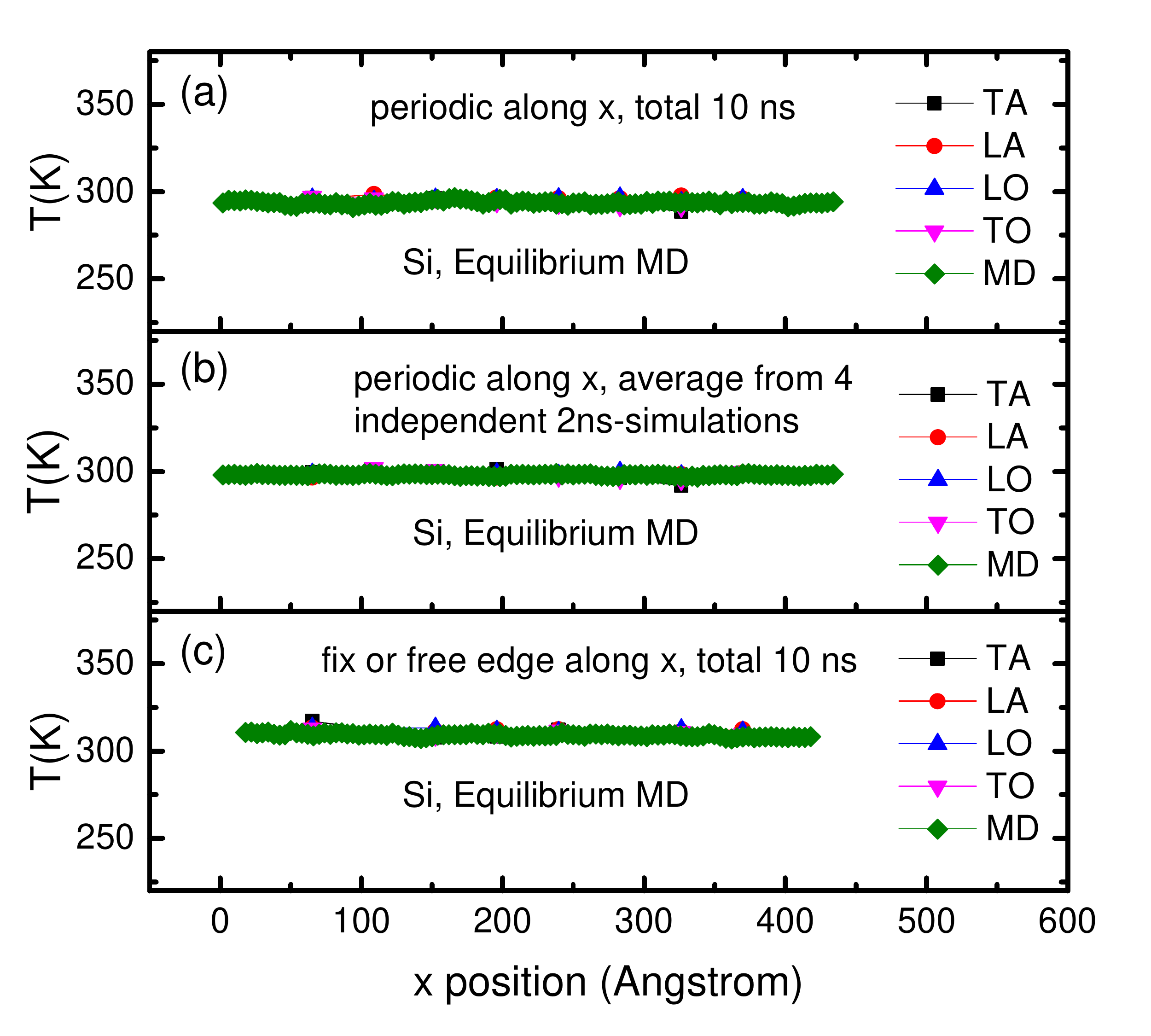}
	\caption{The branch temperatures compared to the MD temperature in equilibrium MD of Si. To benchmark our method, we studied the long-time simulation (a), the ensemble average of short-time simulations (b), and different boundary conditions (c). It can be found that the modal temperatures agree well with the MD temperatures for all the cases, as expected.}\label{suppl_Si_EMD}
\end{figure}

\begin{figure}[!h]
	\centering
	\includegraphics[width= 3.5in]{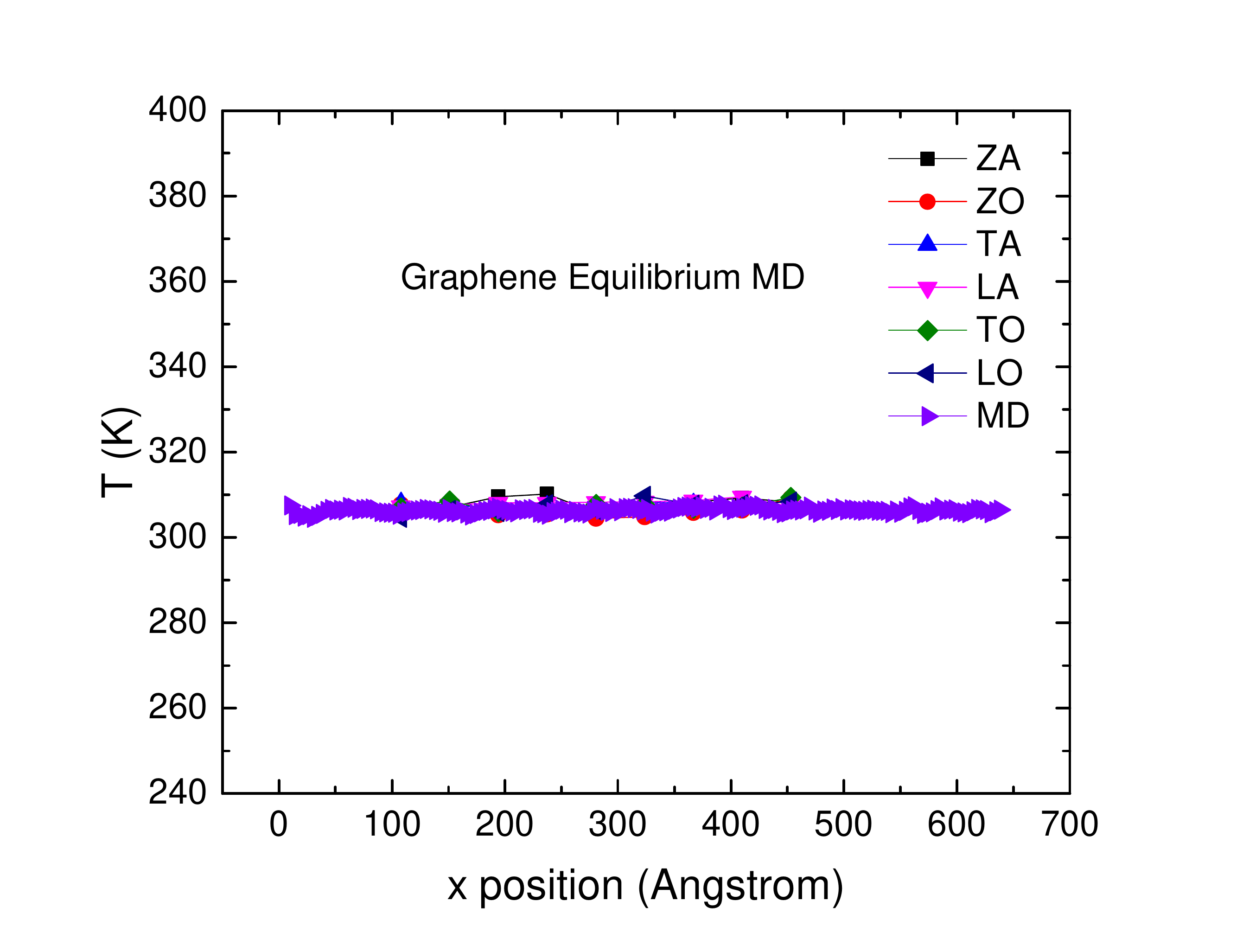}
	\caption{The branch temperatures compared to the MD temperature in equilibrium MD of graphene. They agree well with each other.}\label{fig_argon_comp}
\end{figure}

\section{Formalism of the phonon temperature averaging}
\label{appendB}
\setcounter{figure}{0}

The kinetic energy of the atoms is the $3N$ degrees of freedom times the average energy each degree of freedom $k_BT/2$
\begin{equation}
\label{E_totalatom}
E_{K,atom}=3Nk_BT/2.
\end{equation}
The kinetic energy of the 3N phonon modes is
\begin{equation}
\label{E_totalphonon}
E_{K,phonon}=\frac{1}{2}\sum_{\lambda}^{3N} n_\lambda\hbar\omega_\lambda=\frac{1}{2} \sum_{\lambda}^{3N} E_\lambda = \frac{1}{2}\sum_{\lambda}^{3N} K_BT_\lambda.
\end{equation}
Since the two total energies should be equal to each other, the averaged phonon temperature by combining Eqs.\,(\ref{E_totalatom}) and (\ref{E_totalphonon}) is
\begin{gather}
\bar{T}_\lambda=\frac{1}{3N}\sum_\lambda^{3N} T_\lambda= \frac{1}{3n}\sum_\nu^{3n} \bar{T}_\nu,\\
\bar{T}_\nu=\frac{1}{N_c}\sum_\mathbf{k}^{N_c} T_{\mathbf{k},\nu}=\frac{\int T_{\mathbf{k},\nu} d \mathbf{k}}{\int d \mathbf{k}}.
\end{gather}
Since practically only finite number of phonon modes are calculated, and thus the integral is evaluated by a discrete form:
\begin{gather}
\bar{T}_\nu= \frac{\int k^2 T_{\mathbf{k},\nu} dk}{\int k^2 dk} \approx \frac{\sum_m k_m^2 T_{\mathbf{k},\nu}}{k_m^2},\ \ 3D \\
\bar{T}_\nu= \frac{\int k T_{\mathbf{k},\nu} dk}{\int k dk} \approx \frac{\sum_m k_m T_{\mathbf{k},\nu}}{k_m}. \ \ 2D
\end{gather}

\section{Phonon mean free path spectra obtained by spectral energy density analysis}
\label{appendC}
\setcounter{figure}{0}

The spectral phonon mean free path is obtained by $\Lambda_\lambda=v_\lambda\tau_\lambda$, where $v_\lambda$ is the group velocity, and $\tau_\lambda$ is the phonon relaxation time. $\tau_\lambda$ is obtained by performing the following NMA\,\cite{Dove_book,Qiu2011arXiv, Feng2014Jn,Feng2015jap} based on equilibrium MD simulations,
\begin{equation}
E_\lambda(\omega) = \left| \mathcal{F} [\dot{Q}_\lambda(t)] \right|^2= \frac{C_\lambda}{(\omega-\omega_\lambda^A)^2+(\tau_{\lambda}^{-1})^2/4}. \label{SEDfunction} 
\end{equation}
$\mathcal{F}$ denotes the Fourier transformation. The spectral energy density $E_\lambda(\omega)$ of the phonon mode $\lambda$ is obtained by substituting $\dot{u}_\alpha^{l,b}(t)$ extracted from MD trajectory into Eq.\,(\ref{SEDfunction}), where $C_\lambda$ is a constant for a given $\lambda$. By fitting the spectral energy density as a Lorentzian function, the peak position $\omega_\lambda^A$ and full linewidth $\tau_{NMA,\lambda}^{-1}$ at half maximum are obtained. Our former work\,\cite{Feng2015jap} has shown that Eq.\,(\ref{SEDfunction}) are equivalent to another version of frequency-domain NMA that does not include phonon eigenvectors\,\cite{Koker2009prl,Thomas2010prb}
\begin{align}
\label{Phi2}
&\Phi(\mathbf{k},\omega)\! =\sum_\nu^{3n} E_\lambda(\omega)\\ \nonumber
&=\! \frac{1}{4\pi t_0}\sum_{\alpha}^{3}\sum_{b}^{n}\frac{m_b}{N_c}\left|\sum_l^{N_c}\! \int_0^{t_0}\! \dot{u}_\alpha^{l,b}(t) \exp(i\mathbf{k}\! \cdot\! \mathbf{r}_0^l \!-\! i\omega t)dt \right |^2.
\end{align}
A full discussion about the methods of predicting phonon relaxation time was given in Ref.\,\cite{Feng2014Jn}.

\begin{figure}[tpbh]
	\centering
	\includegraphics[width= 3.4in, height=2.3in]{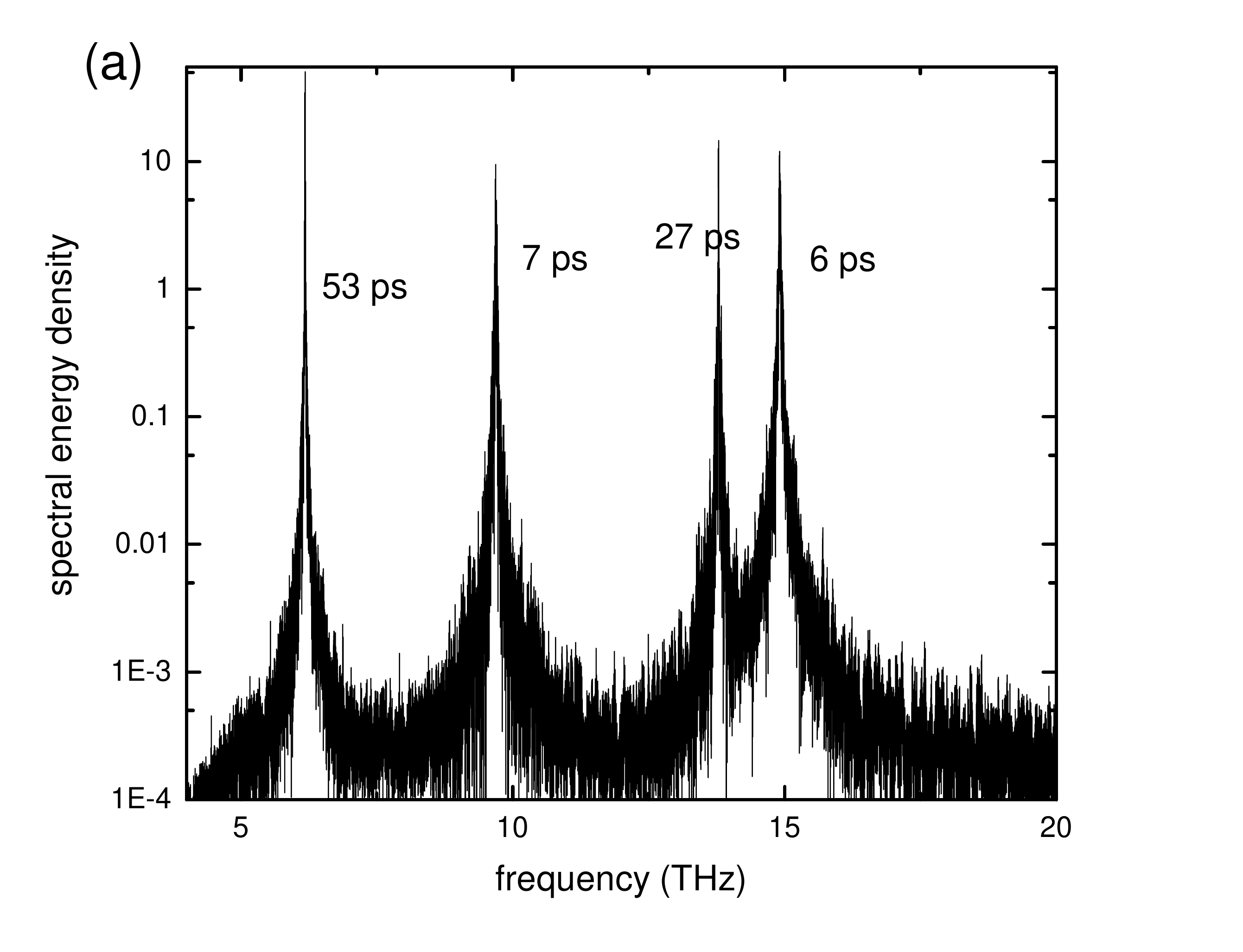}
	\includegraphics[width= 3.4in, height=2.3in]{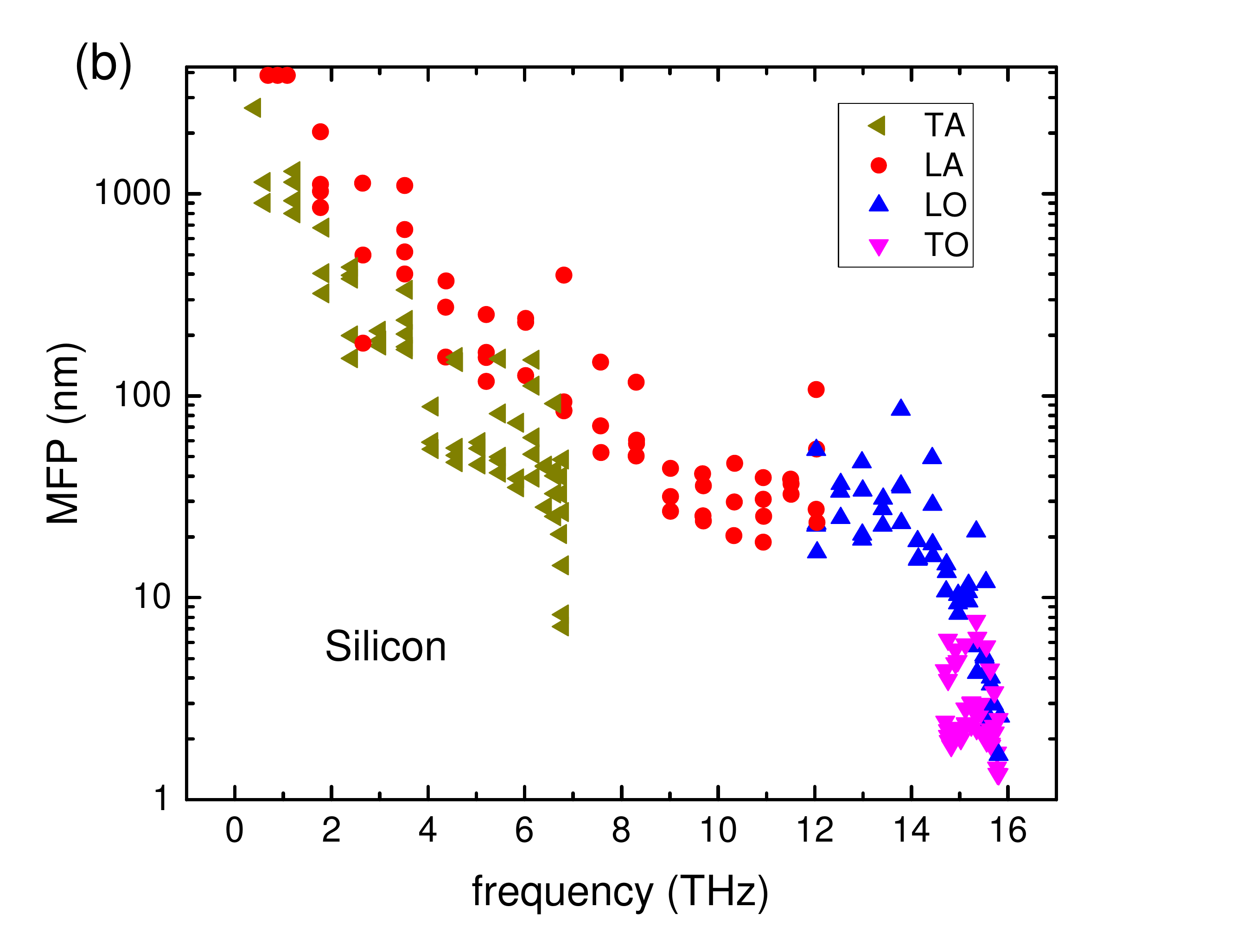}
	\includegraphics[width= 3.4in, height=2.3in]{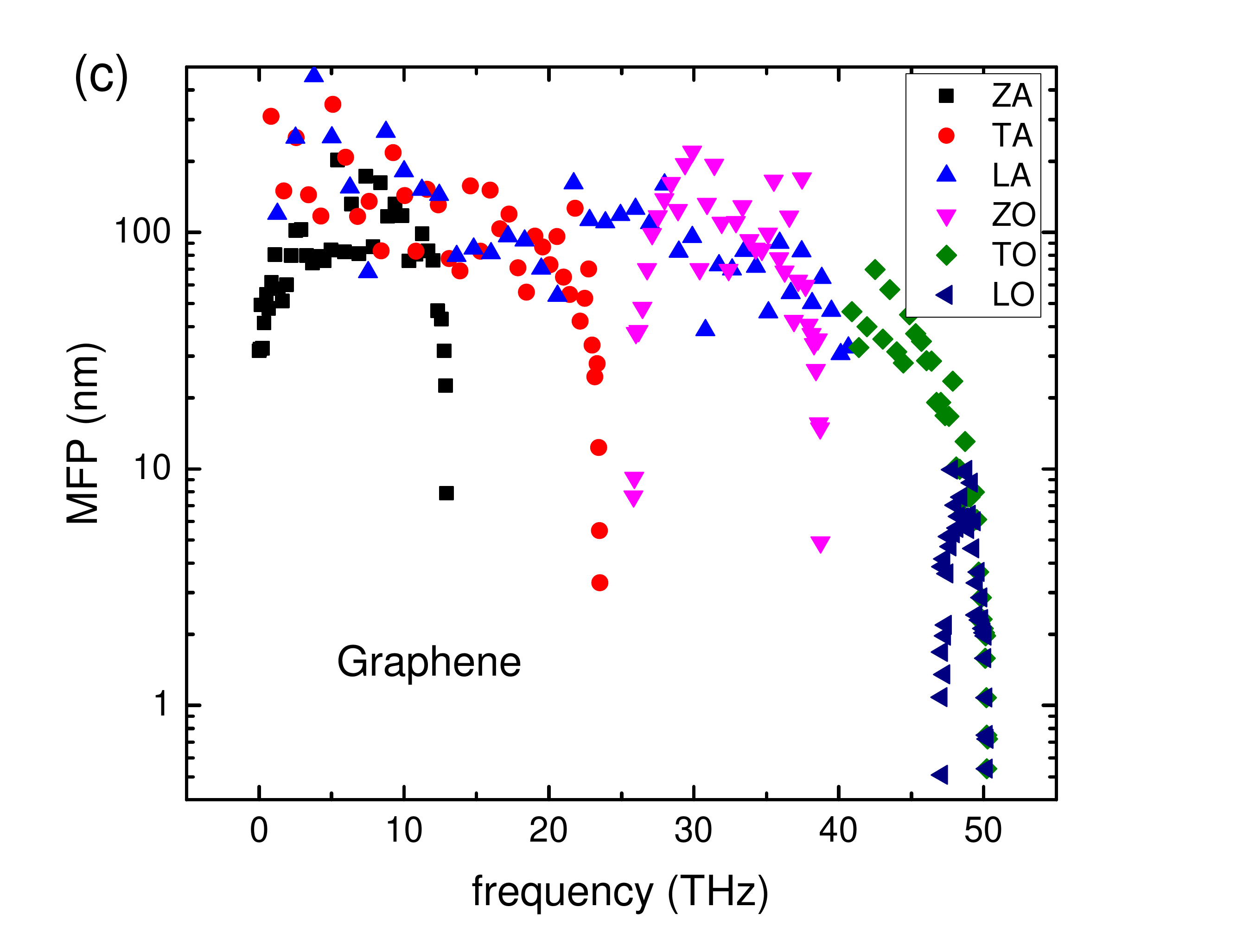}
	\caption{(a) A sample plot showing the process of obtaining the phonon relaxation time from the spectral energy density (SED) in silicon. (b), (c) The spectral phonon MFP in silicon and graphene at room temperature obtained by the SED method.}\label{fig_SED}
\end{figure}

Figure \ref{fig_SED} (a) shows a sample plot of the spectral energy density of Si. The position of the peaks indicate phonon frequencies $\omega_\lambda^A$, and the full width at the half maximum (FWHM) indicates the reciprocal of relaxation time $\tau_{\lambda}^{-1}$. The Figs.\,\ref{fig_SED} (b) and (c) are the phonon MFP as a function of frequency in silicon and graphene, respectively.

\section{Spectral Phonon Temperature of Silicon in Non-equilibrium molecular dynamics under Berendsen thermostat}
\label{appendD}
\setcounter{figure}{0}

\begin{figure}[hbtp]
	\centering
	\includegraphics[width= 3.7in]{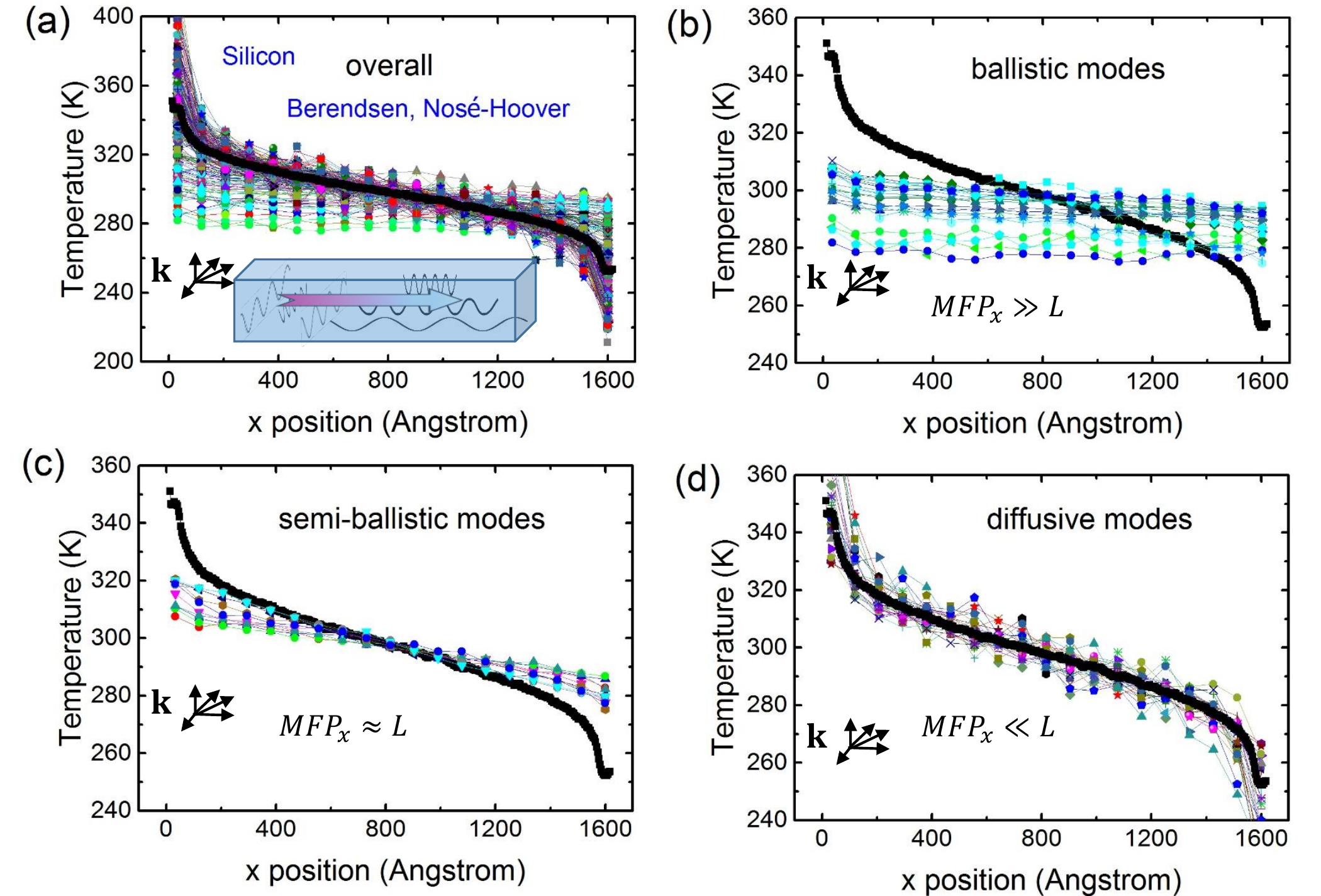}
	\caption{The phonon modal temperatures as a function of position along the $x$ direction in silicon with the Berendsen thermostat. The MD temperature (black) is plotted as a reference. Panel (a) shows the temperatures of all the 240 phonon modes studied in our work. The 240 modes contains multiple directions: [1,0,0], [1,1,0], [1,1,1], [0,1,0], and [0,0,1], while the heat flow is along the [1,0,0] direction. Panels (b), (c), and (d) show some of the 240 modes with flat, slightly steeper, and steep T profiles, respectively.}\label{fig_T_Si_all_Berendsen}
\end{figure}

\begin{figure}[htpb]
	\centering
	\includegraphics[width= 3.7in]{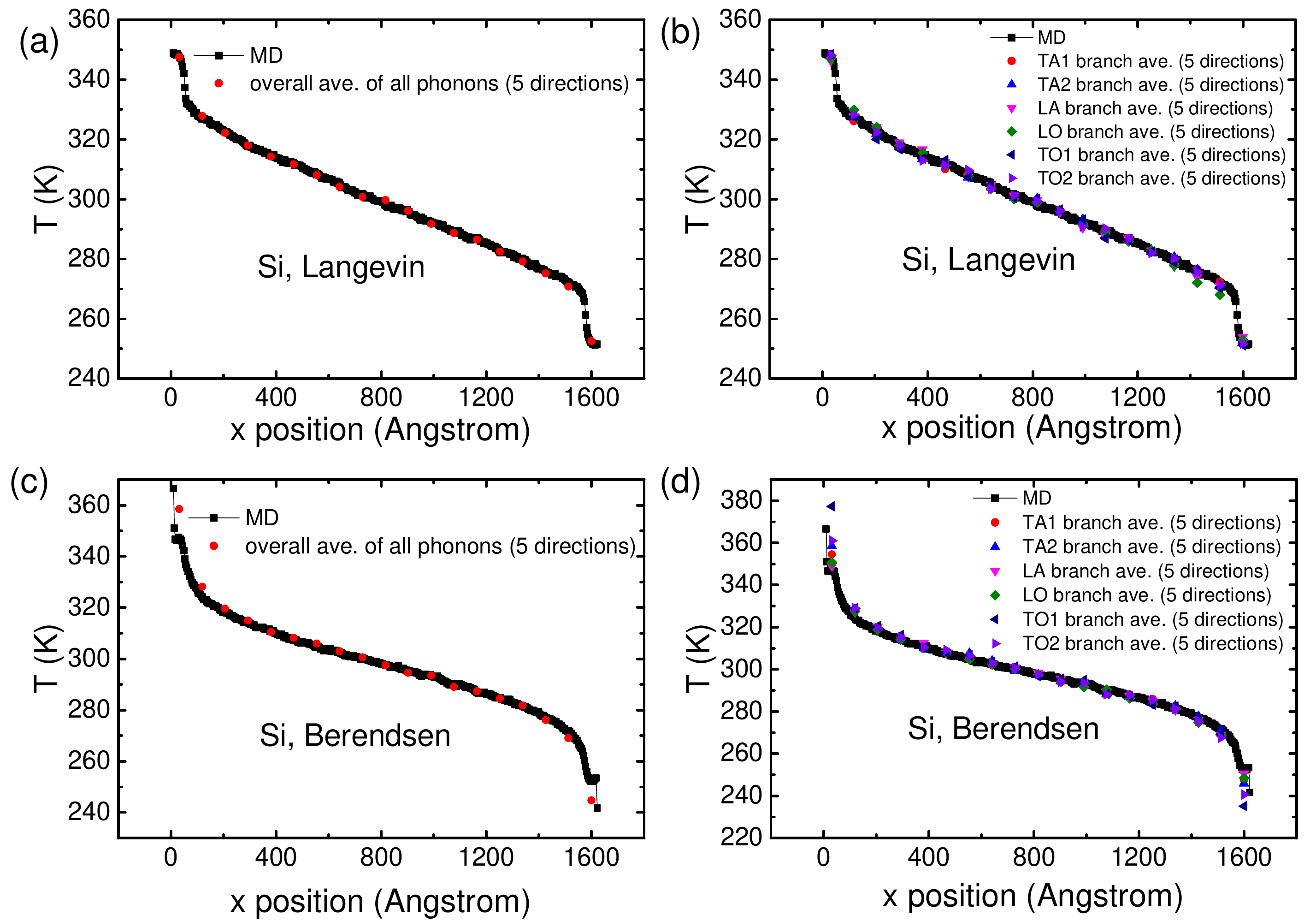}
	\caption{Overall average and branch average phonon temperatures as compared to the MD temperature in silicon with Berendsen (a, b) and Langevin (c, d) thermostats. Panels (a) and (c) show the overall average phonon temperature, while (b) and (c) show the branch temperatures.}\label{fig_average_Si_Tall_Ber_Lang}
\end{figure}

\section{Spectral Phonon Temperature of Graphene under Langevin thermostat}
\label{appendE}
\setcounter{figure}{0}

\begin{figure}[h]
	\centering
	\includegraphics[width= 3.5in]{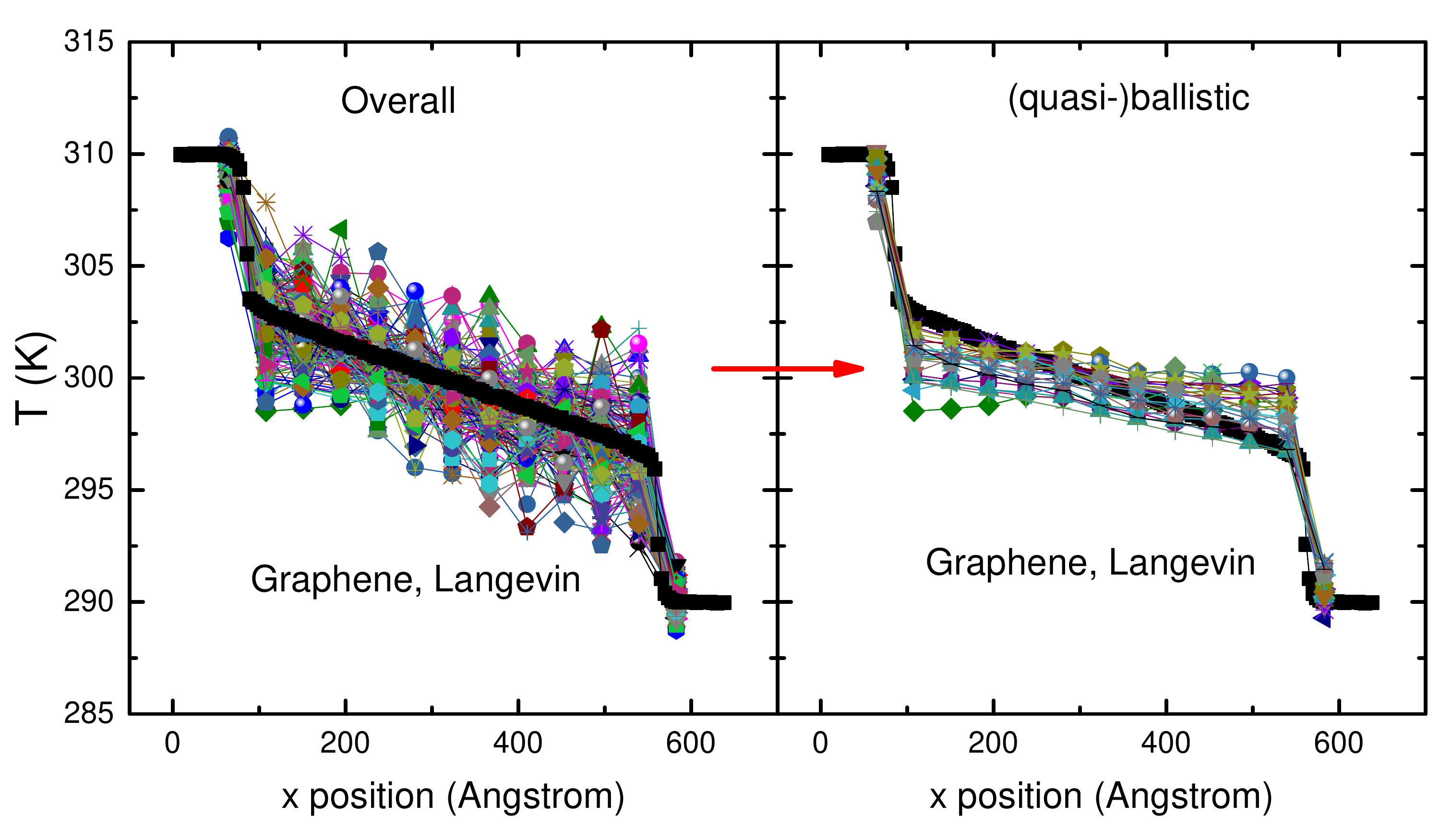}
	\caption{The spectral phonon temperatures as a function of position along the $x$ direction in graphene with the Langevin thermostat. The MD temperature (black) is plotted as a reference. Panel (a) shows the temperatures of all the 240 phonon modes studied in our work. The 240 modes contains multiple directions: $\Gamma-M$, $\Gamma-K$, $\Gamma-M'$, and $\Gamma-K'$, while the heat flow is along the $\Gamma-M$ direction. Panels (b) show some of the 240 modes with flat T profiles. }\label{fig_Tall_graphene_langevin}
\end{figure}

\begin{figure}[h]
	\centering
	\includegraphics[width= 3.5in]{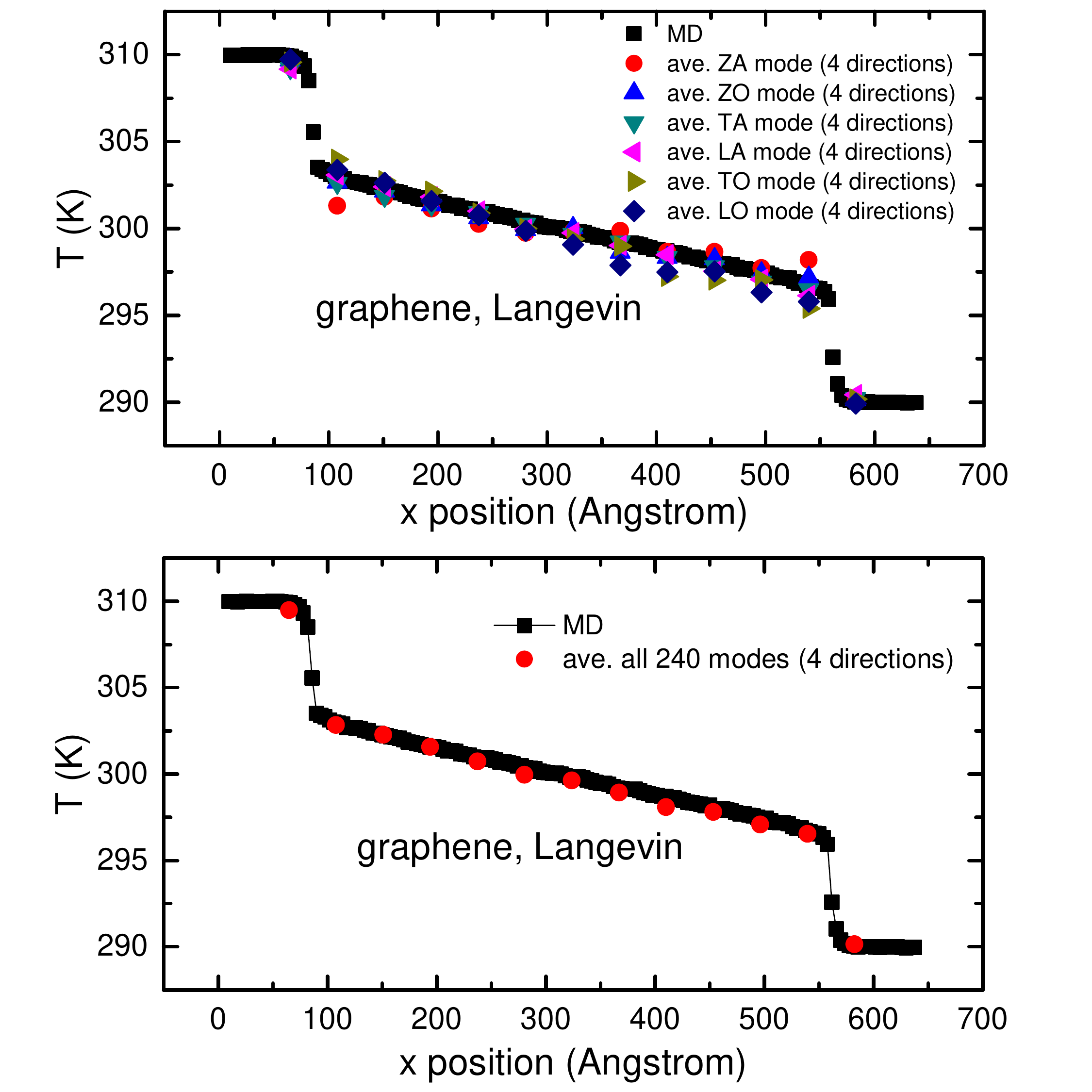}
	\caption{The branch temperature and the average phonon temperature with comparison to the MD temperature in graphene with the Langevin thermostat.}\label{fig_average_graphene_Lang}
\end{figure}

\begin{figure}[h]
	\centering
	\includegraphics[width= 3.5in]{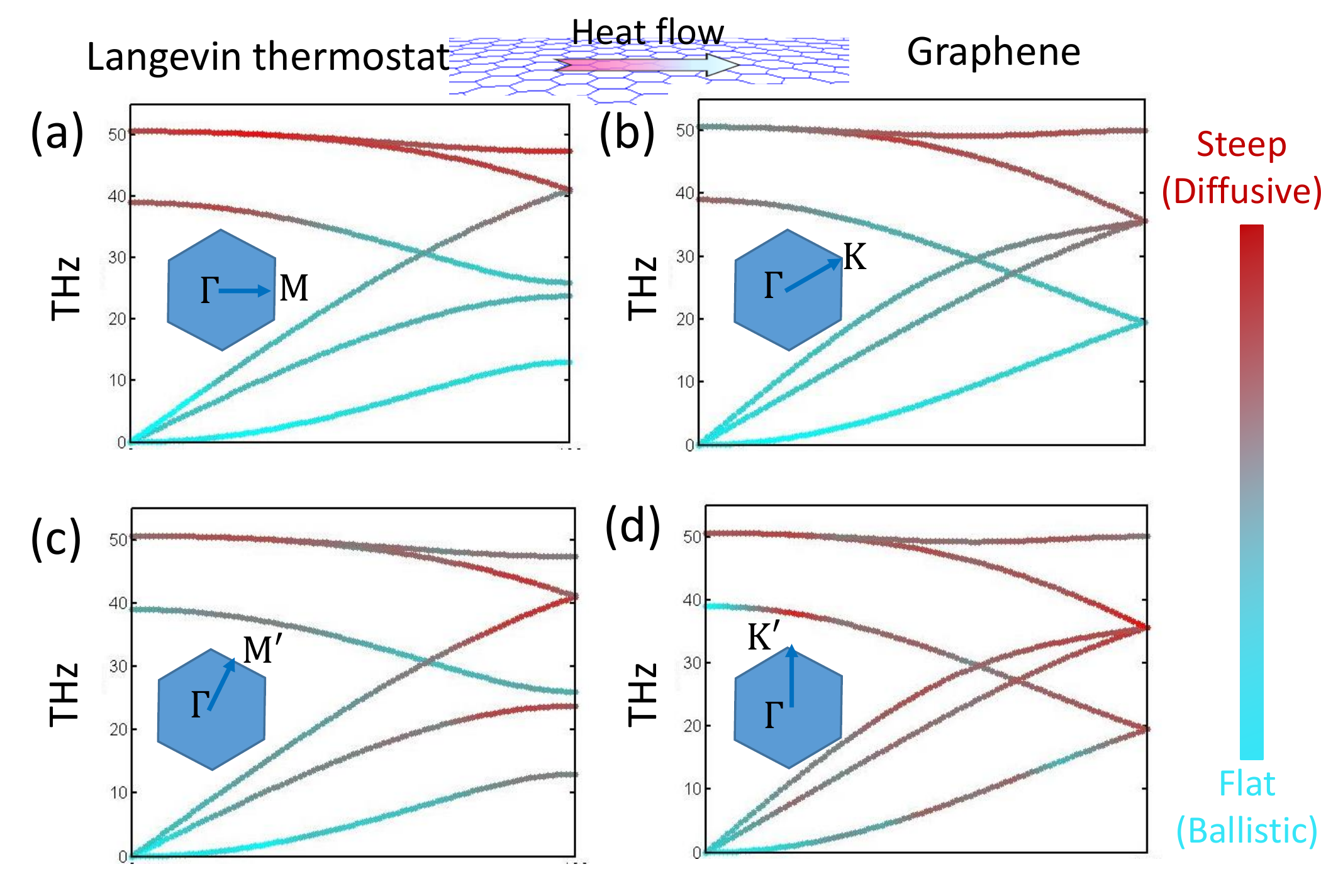}
	\caption{The spectral phonon T gradient represented by color in graphene with Langevin thermostat. Green color represents small T gradient. Red color represents large T gradient.}\label{fig_color_graphene_Langevin}
\end{figure}
\bibliography{bibfile}

\begin{thebibliography}{61}
\expandafter\ifx\csname natexlab\endcsname\relax\def\natexlab#1{#1}\fi
\expandafter\ifx\csname bibnamefont\endcsname\relax
  \def\bibnamefont#1{#1}\fi
\expandafter\ifx\csname bibfnamefont\endcsname\relax
  \def\bibfnamefont#1{#1}\fi
\expandafter\ifx\csname citenamefont\endcsname\relax
  \def\citenamefont#1{#1}\fi
\expandafter\ifx\csname url\endcsname\relax
  \def\url#1{\texttt{#1}}\fi
\expandafter\ifx\csname urlprefix\endcsname\relax\def\urlprefix{URL }\fi
\providecommand{\bibinfo}[2]{#2}
\providecommand{\eprint}[2][]{\url{#2}}

\bibitem[{\citenamefont{Minnich}(2012)}]{Minnich2012}
\bibinfo{author}{\bibfnamefont{a.~J.} \bibnamefont{Minnich}},
  \bibinfo{journal}{Physical Review Letters} \textbf{\bibinfo{volume}{109}},
  \bibinfo{pages}{205901} (\bibinfo{year}{2012}), ISSN
  \bibinfo{issn}{0031-9007},
  \urlprefix\url{http://link.aps.org/doi/10.1103/PhysRevLett.109.205901}.

\bibitem[{\citenamefont{Regner et~al.}(2013)\citenamefont{Regner, Sellan, Su,
  Amon, McGaughey, and Malen}}]{Regner2013nc}
\bibinfo{author}{\bibfnamefont{K.~T.} \bibnamefont{Regner}},
  \bibinfo{author}{\bibfnamefont{D.~P.} \bibnamefont{Sellan}},
  \bibinfo{author}{\bibfnamefont{Z.}~\bibnamefont{Su}},
  \bibinfo{author}{\bibfnamefont{C.~H.} \bibnamefont{Amon}},
  \bibinfo{author}{\bibfnamefont{A.~J.~H.} \bibnamefont{McGaughey}},
  \bibnamefont{and} \bibinfo{author}{\bibfnamefont{J.~a.} \bibnamefont{Malen}},
  \bibinfo{journal}{Nature communications} \textbf{\bibinfo{volume}{4}},
  \bibinfo{pages}{1640} (\bibinfo{year}{2013}), ISSN \bibinfo{issn}{2041-1723},
  \urlprefix\url{http://www.ncbi.nlm.nih.gov/pubmed/23535661}.

\bibitem[{\citenamefont{Hu et~al.}(2015)\citenamefont{Hu, Zeng, Minnich, and
  {Dresselhaus, Mildred S. Chen}}}]{Hu2015nn}
\bibinfo{author}{\bibfnamefont{Y.}~\bibnamefont{Hu}},
  \bibinfo{author}{\bibfnamefont{L.}~\bibnamefont{Zeng}},
  \bibinfo{author}{\bibfnamefont{A.~J.} \bibnamefont{Minnich}},
  \bibnamefont{and}
  \bibinfo{author}{\bibfnamefont{G.}~\bibnamefont{{Dresselhaus, Mildred S.
  Chen}}}, \bibinfo{journal}{Nature nanotechnology}
  \textbf{\bibinfo{volume}{10}}, \bibinfo{pages}{701} (\bibinfo{year}{2015}),
  ISSN \bibinfo{issn}{1748-3387},
  \urlprefix\url{http://dx.doi.org/10.1038/nnano.2015.109}.

\bibitem[{\citenamefont{Zeng et~al.}(2015)\citenamefont{Zeng, Collins, Hu,
  Luckyanova, Maznev, Huberman, Chiloyan, Zhou, Huang, Nelson
  et~al.}}]{Zeng2015sr}
\bibinfo{author}{\bibfnamefont{L.}~\bibnamefont{Zeng}},
  \bibinfo{author}{\bibfnamefont{K.~C.} \bibnamefont{Collins}},
  \bibinfo{author}{\bibfnamefont{Y.}~\bibnamefont{Hu}},
  \bibinfo{author}{\bibfnamefont{M.~N.} \bibnamefont{Luckyanova}},
  \bibinfo{author}{\bibfnamefont{A.~A.} \bibnamefont{Maznev}},
  \bibinfo{author}{\bibfnamefont{S.}~\bibnamefont{Huberman}},
  \bibinfo{author}{\bibfnamefont{V.}~\bibnamefont{Chiloyan}},
  \bibinfo{author}{\bibfnamefont{J.}~\bibnamefont{Zhou}},
  \bibinfo{author}{\bibfnamefont{X.}~\bibnamefont{Huang}},
  \bibinfo{author}{\bibfnamefont{K.~A.} \bibnamefont{Nelson}},
  \bibnamefont{et~al.}, \bibinfo{journal}{Scientific Reports}
  \textbf{\bibinfo{volume}{5}}, \bibinfo{pages}{17131} (\bibinfo{year}{2015}),
  ISSN \bibinfo{issn}{2045-2322},
  \urlprefix\url{http://www.nature.com/articles/srep17131}.

\bibitem[{\citenamefont{Cuffe et~al.}(2015)\citenamefont{Cuffe, Eliason,
  Maznev, Collins, Johnson, Shchepetov, Prunnila, Ahopelto, {Sotomayor Torres},
  Chen et~al.}}]{Cuffe2015}
\bibinfo{author}{\bibfnamefont{J.}~\bibnamefont{Cuffe}},
  \bibinfo{author}{\bibfnamefont{J.~K.} \bibnamefont{Eliason}},
  \bibinfo{author}{\bibfnamefont{a.~a.} \bibnamefont{Maznev}},
  \bibinfo{author}{\bibfnamefont{K.~C.} \bibnamefont{Collins}},
  \bibinfo{author}{\bibfnamefont{J.~a.} \bibnamefont{Johnson}},
  \bibinfo{author}{\bibfnamefont{A.}~\bibnamefont{Shchepetov}},
  \bibinfo{author}{\bibfnamefont{M.}~\bibnamefont{Prunnila}},
  \bibinfo{author}{\bibfnamefont{J.}~\bibnamefont{Ahopelto}},
  \bibinfo{author}{\bibfnamefont{C.~M.} \bibnamefont{{Sotomayor Torres}}},
  \bibinfo{author}{\bibfnamefont{G.}~\bibnamefont{Chen}}, \bibnamefont{et~al.},
  \bibinfo{journal}{Physical Review B} \textbf{\bibinfo{volume}{91}},
  \bibinfo{pages}{245423} (\bibinfo{year}{2015}), ISSN
  \bibinfo{issn}{1098-0121},
  \urlprefix\url{http://link.aps.org/doi/10.1103/PhysRevB.91.245423}.

\bibitem[{\citenamefont{Lee et~al.}(2015)\citenamefont{Lee, Lim, and
  Yang}}]{Lee2015nl}
\bibinfo{author}{\bibfnamefont{J.}~\bibnamefont{Lee}},
  \bibinfo{author}{\bibfnamefont{J.}~\bibnamefont{Lim}}, \bibnamefont{and}
  \bibinfo{author}{\bibfnamefont{P.}~\bibnamefont{Yang}},
  \bibinfo{journal}{Nano Letters} \textbf{\bibinfo{volume}{15}},
  \bibinfo{pages}{3273−3279} (\bibinfo{year}{2015}), ISSN
  \bibinfo{issn}{1530-6984},
  \urlprefix\url{http://pubs.acs.org/doi/abs/10.1021/acs.nanolett.5b00495}.

\bibitem[{\citenamefont{Zen et~al.}(2014)\citenamefont{Zen, Puurtinen, Isotalo,
  Chaudhuri, and Maasilta}}]{Zen2014nc}
\bibinfo{author}{\bibfnamefont{N.}~\bibnamefont{Zen}},
  \bibinfo{author}{\bibfnamefont{T.~a.} \bibnamefont{Puurtinen}},
  \bibinfo{author}{\bibfnamefont{T.~J.} \bibnamefont{Isotalo}},
  \bibinfo{author}{\bibfnamefont{S.}~\bibnamefont{Chaudhuri}},
  \bibnamefont{and} \bibinfo{author}{\bibfnamefont{I.~J.}
  \bibnamefont{Maasilta}}, \bibinfo{journal}{Nature Communications}
  \textbf{\bibinfo{volume}{5}}, \bibinfo{pages}{3435} (\bibinfo{year}{2014}),
  ISSN \bibinfo{issn}{2041-1723},
  \urlprefix\url{http://www.nature.com/doifinder/10.1038/ncomms4435}.

\bibitem[{\citenamefont{Maldovan}(2015)}]{Maldovan2015nm}
\bibinfo{author}{\bibfnamefont{M.}~\bibnamefont{Maldovan}},
  \bibinfo{journal}{Nature Materials} \textbf{\bibinfo{volume}{14}},
  \bibinfo{pages}{667} (\bibinfo{year}{2015}), ISSN \bibinfo{issn}{1476-1122},
  \urlprefix\url{http://www.nature.com/nmat/journal/v14/n7/full/nmat4308.html?WT.ec{\_}id=NMAT-201507}.

\bibitem[{\citenamefont{Ravichandran et~al.}(2013)\citenamefont{Ravichandran,
  Yadav, Cheaito, Rossen, Soukiassian, Suresha, Duda, Foley, Lee, Zhu
  et~al.}}]{Ravichandran2013nm}
\bibinfo{author}{\bibfnamefont{J.}~\bibnamefont{Ravichandran}},
  \bibinfo{author}{\bibfnamefont{A.~K.} \bibnamefont{Yadav}},
  \bibinfo{author}{\bibfnamefont{R.}~\bibnamefont{Cheaito}},
  \bibinfo{author}{\bibfnamefont{P.~B.} \bibnamefont{Rossen}},
  \bibinfo{author}{\bibfnamefont{A.}~\bibnamefont{Soukiassian}},
  \bibinfo{author}{\bibfnamefont{S.~J.} \bibnamefont{Suresha}},
  \bibinfo{author}{\bibfnamefont{J.~C.} \bibnamefont{Duda}},
  \bibinfo{author}{\bibfnamefont{B.~M.} \bibnamefont{Foley}},
  \bibinfo{author}{\bibfnamefont{C.-H.} \bibnamefont{Lee}},
  \bibinfo{author}{\bibfnamefont{Y.}~\bibnamefont{Zhu}}, \bibnamefont{et~al.},
  \bibinfo{journal}{Nature materials} \textbf{\bibinfo{volume}{13}},
  \bibinfo{pages}{168} (\bibinfo{year}{2013}), ISSN \bibinfo{issn}{1476-1122},
  \urlprefix\url{http://www.ncbi.nlm.nih.gov/pubmed/24317186}.

\bibitem[{\citenamefont{Luckyanova et~al.}(2012)\citenamefont{Luckyanova, Garg,
  Esfarjani, Jandl, Bulsara, Schmidt, Minnich, Chen, Dresselhaus, Ren
  et~al.}}]{luckyanova2012coherent}
\bibinfo{author}{\bibfnamefont{M.~N.} \bibnamefont{Luckyanova}},
  \bibinfo{author}{\bibfnamefont{J.}~\bibnamefont{Garg}},
  \bibinfo{author}{\bibfnamefont{K.}~\bibnamefont{Esfarjani}},
  \bibinfo{author}{\bibfnamefont{A.}~\bibnamefont{Jandl}},
  \bibinfo{author}{\bibfnamefont{M.~T.} \bibnamefont{Bulsara}},
  \bibinfo{author}{\bibfnamefont{A.~J.} \bibnamefont{Schmidt}},
  \bibinfo{author}{\bibfnamefont{A.~J.} \bibnamefont{Minnich}},
  \bibinfo{author}{\bibfnamefont{S.}~\bibnamefont{Chen}},
  \bibinfo{author}{\bibfnamefont{M.~S.} \bibnamefont{Dresselhaus}},
  \bibinfo{author}{\bibfnamefont{Z.}~\bibnamefont{Ren}}, \bibnamefont{et~al.},
  \bibinfo{journal}{Science} \textbf{\bibinfo{volume}{338}},
  \bibinfo{pages}{936} (\bibinfo{year}{2012}), ISSN \bibinfo{issn}{0163-1829},
  \urlprefix\url{http://www.sciencemag.org/content/338/6109/936.short}.

\bibitem[{\citenamefont{Hsiao et~al.}(2013)\citenamefont{Hsiao, Chang, Liou,
  Chu, Lee, and Chang}}]{Hsiao2013nn}
\bibinfo{author}{\bibfnamefont{T.-K.} \bibnamefont{Hsiao}},
  \bibinfo{author}{\bibfnamefont{H.-K.} \bibnamefont{Chang}},
  \bibinfo{author}{\bibfnamefont{S.-C.} \bibnamefont{Liou}},
  \bibinfo{author}{\bibfnamefont{M.-W.} \bibnamefont{Chu}},
  \bibinfo{author}{\bibfnamefont{S.-C.} \bibnamefont{Lee}}, \bibnamefont{and}
  \bibinfo{author}{\bibfnamefont{C.-W.} \bibnamefont{Chang}},
  \bibinfo{journal}{Nature nanotechnology} \textbf{\bibinfo{volume}{8}},
  \bibinfo{pages}{534} (\bibinfo{year}{2013}), ISSN \bibinfo{issn}{1748-3395},
  \urlprefix\url{http://www.ncbi.nlm.nih.gov/pubmed/23812186}.

\bibitem[{\citenamefont{Feng and Ruan}(2014)}]{Feng2014Jn}
\bibinfo{author}{\bibfnamefont{T.}~\bibnamefont{Feng}} \bibnamefont{and}
  \bibinfo{author}{\bibfnamefont{X.}~\bibnamefont{Ruan}},
  \bibinfo{journal}{Journal of Nanomaterials} \textbf{\bibinfo{volume}{2014}},
  \bibinfo{pages}{206370} (\bibinfo{year}{2014}), ISSN
  \bibinfo{issn}{1687-4110},
  \urlprefix\url{http://www.hindawi.com/journals/jnm/2014/206370/}.

\bibitem[{\citenamefont{Minnich}(2015)}]{Minnich2015prb}
\bibinfo{author}{\bibfnamefont{A.~J.} \bibnamefont{Minnich}},
  \bibinfo{journal}{Phys. Rev. B} \textbf{\bibinfo{volume}{92}},
  \bibinfo{pages}{085203} (\bibinfo{year}{2015}),
  \urlprefix\url{http://link.aps.org/doi/10.1103/PhysRevB.92.085203}.

\bibitem[{\citenamefont{Tian et~al.}(2012)\citenamefont{Tian, Esfarjani, and
  Chen}}]{Tian2012Enhancing}
\bibinfo{author}{\bibfnamefont{Z.}~\bibnamefont{Tian}},
  \bibinfo{author}{\bibfnamefont{K.}~\bibnamefont{Esfarjani}},
  \bibnamefont{and} \bibinfo{author}{\bibfnamefont{G.}~\bibnamefont{Chen}},
  \bibinfo{journal}{Phys. Rev. B} \textbf{\bibinfo{volume}{86}},
  \bibinfo{pages}{235304} (\bibinfo{year}{2012}),
  \urlprefix\url{http://link.aps.org/doi/10.1103/PhysRevB.86.235304}.

\bibitem[{\citenamefont{Dechaumphai et~al.}(2014)\citenamefont{Dechaumphai, Lu,
  Kan, Moon, Fullerton, Liu, and Chen}}]{Dec2014NL}
\bibinfo{author}{\bibfnamefont{E.}~\bibnamefont{Dechaumphai}},
  \bibinfo{author}{\bibfnamefont{D.}~\bibnamefont{Lu}},
  \bibinfo{author}{\bibfnamefont{J.~J.} \bibnamefont{Kan}},
  \bibinfo{author}{\bibfnamefont{J.}~\bibnamefont{Moon}},
  \bibinfo{author}{\bibfnamefont{E.~E.} \bibnamefont{Fullerton}},
  \bibinfo{author}{\bibfnamefont{Z.}~\bibnamefont{Liu}}, \bibnamefont{and}
  \bibinfo{author}{\bibfnamefont{R.}~\bibnamefont{Chen}},
  \bibinfo{journal}{Nano letters} \textbf{\bibinfo{volume}{14}},
  \bibinfo{pages}{2448} (\bibinfo{year}{2014}), ISSN \bibinfo{issn}{1530-6992},
  \urlprefix\url{http://www.ncbi.nlm.nih.gov/pubmed/24730544}.

\bibitem[{\citenamefont{Seol et~al.}(2010)\citenamefont{Seol, Jo, Moore,
  Lindsay, Aitken, Pettes, Li, Yao, Huang, Broido et~al.}}]{Seol2010Science}
\bibinfo{author}{\bibfnamefont{J.~H.} \bibnamefont{Seol}},
  \bibinfo{author}{\bibfnamefont{I.}~\bibnamefont{Jo}},
  \bibinfo{author}{\bibfnamefont{A.~L.} \bibnamefont{Moore}},
  \bibinfo{author}{\bibfnamefont{L.}~\bibnamefont{Lindsay}},
  \bibinfo{author}{\bibfnamefont{Z.~H.} \bibnamefont{Aitken}},
  \bibinfo{author}{\bibfnamefont{M.~T.} \bibnamefont{Pettes}},
  \bibinfo{author}{\bibfnamefont{X.}~\bibnamefont{Li}},
  \bibinfo{author}{\bibfnamefont{Z.}~\bibnamefont{Yao}},
  \bibinfo{author}{\bibfnamefont{R.}~\bibnamefont{Huang}},
  \bibinfo{author}{\bibfnamefont{D.}~\bibnamefont{Broido}},
  \bibnamefont{et~al.}, \bibinfo{journal}{Science}
  \textbf{\bibinfo{volume}{328}}, \bibinfo{pages}{213} (\bibinfo{year}{2010}),
  ISSN \bibinfo{issn}{0036-8075},
  \urlprefix\url{http://www.sciencemag.org/content/328/5975/213.short
  http://www.sciencemag.org/content/328/5975/213.abstract}.

\bibitem[{\citenamefont{Chen et~al.}(2014)\citenamefont{Chen, Mante, Chang,
  Yang, Chen, Huang, Chen, Chen, Gusev, and Sun}}]{Chen2014NL}
\bibinfo{author}{\bibfnamefont{I.-J.} \bibnamefont{Chen}},
  \bibinfo{author}{\bibfnamefont{P.-A.} \bibnamefont{Mante}},
  \bibinfo{author}{\bibfnamefont{C.-K.} \bibnamefont{Chang}},
  \bibinfo{author}{\bibfnamefont{S.-C.} \bibnamefont{Yang}},
  \bibinfo{author}{\bibfnamefont{H.-Y.} \bibnamefont{Chen}},
  \bibinfo{author}{\bibfnamefont{Y.-R.} \bibnamefont{Huang}},
  \bibinfo{author}{\bibfnamefont{L.-C.} \bibnamefont{Chen}},
  \bibinfo{author}{\bibfnamefont{K.-H.} \bibnamefont{Chen}},
  \bibinfo{author}{\bibfnamefont{V.}~\bibnamefont{Gusev}}, \bibnamefont{and}
  \bibinfo{author}{\bibfnamefont{C.-K.} \bibnamefont{Sun}},
  \bibinfo{journal}{Nano letters} \textbf{\bibinfo{volume}{14}},
  \bibinfo{pages}{1317} (\bibinfo{year}{2014}).

\bibitem[{\citenamefont{Li and Yang}(2012)}]{Li2012Effect}
\bibinfo{author}{\bibfnamefont{X.}~\bibnamefont{Li}} \bibnamefont{and}
  \bibinfo{author}{\bibfnamefont{R.}~\bibnamefont{Yang}},
  \bibinfo{journal}{Physical Review B} \textbf{\bibinfo{volume}{86}},
  \bibinfo{pages}{054305} (\bibinfo{year}{2012}), ISSN
  \bibinfo{issn}{1098-0121},
  \urlprefix\url{http://link.aps.org/doi/10.1103/PhysRevB.86.054305}.

\bibitem[{\citenamefont{Vermeersch et~al.}(2014)\citenamefont{Vermeersch,
  Mohammed, Pernot, Koh, and Shakouri}}]{Verm2014prb}
\bibinfo{author}{\bibfnamefont{B.}~\bibnamefont{Vermeersch}},
  \bibinfo{author}{\bibfnamefont{A.~M.~S.} \bibnamefont{Mohammed}},
  \bibinfo{author}{\bibfnamefont{G.}~\bibnamefont{Pernot}},
  \bibinfo{author}{\bibfnamefont{Y.~R.} \bibnamefont{Koh}}, \bibnamefont{and}
  \bibinfo{author}{\bibfnamefont{A.}~\bibnamefont{Shakouri}},
  \bibinfo{journal}{Physical Review B} \textbf{\bibinfo{volume}{90}},
  \bibinfo{pages}{014306} (\bibinfo{year}{2014}), ISSN
  \bibinfo{issn}{1098-0121},
  \urlprefix\url{http://link.aps.org/doi/10.1103/PhysRevB.90.014306}.

\bibitem[{\citenamefont{Siemens et~al.}(2010)\citenamefont{Siemens, Li, Yang,
  Nelson, Anderson, Murnane, and Kapteyn}}]{Siemens2010nm}
\bibinfo{author}{\bibfnamefont{M.~E.} \bibnamefont{Siemens}},
  \bibinfo{author}{\bibfnamefont{Q.}~\bibnamefont{Li}},
  \bibinfo{author}{\bibfnamefont{R.}~\bibnamefont{Yang}},
  \bibinfo{author}{\bibfnamefont{K.~A.} \bibnamefont{Nelson}},
  \bibinfo{author}{\bibfnamefont{E.~H.} \bibnamefont{Anderson}},
  \bibinfo{author}{\bibfnamefont{M.~M.} \bibnamefont{Murnane}},
  \bibnamefont{and} \bibinfo{author}{\bibfnamefont{H.~C.}
  \bibnamefont{Kapteyn}}, \bibinfo{journal}{Nature Materials}
  \textbf{\bibinfo{volume}{9}}, \bibinfo{pages}{26} (\bibinfo{year}{2010}),
  ISSN \bibinfo{issn}{1476-1122},
  \urlprefix\url{http://www.nature.com/doifinder/10.1038/nmat2568}.

\bibitem[{\citenamefont{Zhou et~al.}(2016)\citenamefont{Zhou, Zhang, and
  Hu}}]{Zhou2016Ns}
\bibinfo{author}{\bibfnamefont{Y.}~\bibnamefont{Zhou}},
  \bibinfo{author}{\bibfnamefont{X.}~\bibnamefont{Zhang}}, \bibnamefont{and}
  \bibinfo{author}{\bibfnamefont{M.}~\bibnamefont{Hu}},
  \bibinfo{journal}{Nanoscale} \textbf{\bibinfo{volume}{8}},
  \bibinfo{pages}{1994} (\bibinfo{year}{2016}),
  \urlprefix\url{http://dx.doi.org/10.1039/C5NR06855J}.

\bibitem[{\citenamefont{Dunn et~al.}(2016)\citenamefont{Dunn, Antillon,
  Maassen, Lundstrom, and Strachan}}]{Dunn2016}
\bibinfo{author}{\bibfnamefont{J.}~\bibnamefont{Dunn}},
  \bibinfo{author}{\bibfnamefont{E.}~\bibnamefont{Antillon}},
  \bibinfo{author}{\bibfnamefont{J.}~\bibnamefont{Maassen}},
  \bibinfo{author}{\bibfnamefont{M.}~\bibnamefont{Lundstrom}},
  \bibnamefont{and} \bibinfo{author}{\bibfnamefont{A.}~\bibnamefont{Strachan}},
  \bibinfo{journal}{Journal of Applied Physics} \textbf{\bibinfo{volume}{120}},
  \bibinfo{pages}{225112} (\bibinfo{year}{2016}), ISSN
  \bibinfo{issn}{0021-8979},
  \urlprefix\url{http://aip.scitation.org/doi/10.1063/1.4971254}.

\bibitem[{\citenamefont{Rurali et~al.}(2016)\citenamefont{Rurali, Colombo,
  Cartoixa, Wilhelmsen, Trinh, Bedeaux, and Kjelstrup}}]{C6CP01872F}
\bibinfo{author}{\bibfnamefont{R.}~\bibnamefont{Rurali}},
  \bibinfo{author}{\bibfnamefont{L.}~\bibnamefont{Colombo}},
  \bibinfo{author}{\bibfnamefont{X.}~\bibnamefont{Cartoixa}},
  \bibinfo{author}{\bibfnamefont{O.}~\bibnamefont{Wilhelmsen}},
  \bibinfo{author}{\bibfnamefont{T.~T.} \bibnamefont{Trinh}},
  \bibinfo{author}{\bibfnamefont{D.}~\bibnamefont{Bedeaux}}, \bibnamefont{and}
  \bibinfo{author}{\bibfnamefont{S.}~\bibnamefont{Kjelstrup}},
  \bibinfo{journal}{Phys. Chem. Chem. Phys.} \textbf{\bibinfo{volume}{18}},
  \bibinfo{pages}{13741} (\bibinfo{year}{2016}),
  \urlprefix\url{http://dx.doi.org/10.1039/C6CP01872F}.

\bibitem[{\citenamefont{Dove}(1993)}]{Dove_book}
\bibinfo{author}{\bibfnamefont{M.~T.} \bibnamefont{Dove}},
  \emph{\bibinfo{title}{{Introduction to Lattice Dynamics}}}
  (\bibinfo{publisher}{Cambridge University Press}, \bibinfo{address}{New York,
  USA}, \bibinfo{year}{1993}).

\bibitem[{\citenamefont{Zhou and Hu}(2015)}]{Zhou2015prb1}
\bibinfo{author}{\bibfnamefont{Y.}~\bibnamefont{Zhou}} \bibnamefont{and}
  \bibinfo{author}{\bibfnamefont{M.}~\bibnamefont{Hu}},
  \bibinfo{journal}{Physical Review B} \textbf{\bibinfo{volume}{92}},
  \bibinfo{pages}{195204} (\bibinfo{year}{2015}), ISSN
  \bibinfo{issn}{1098-0121},
  \urlprefix\url{http://journals.aps.org/prb/abstract/10.1103/PhysRevB.92.195204}.

\bibitem[{\citenamefont{Mann et~al.}(2006)\citenamefont{Mann, Pop, Cao, Wang,
  Goodson, and Dai}}]{Mann2006}
\bibinfo{author}{\bibfnamefont{D.}~\bibnamefont{Mann}},
  \bibinfo{author}{\bibfnamefont{E.}~\bibnamefont{Pop}},
  \bibinfo{author}{\bibfnamefont{J.}~\bibnamefont{Cao}},
  \bibinfo{author}{\bibfnamefont{Q.}~\bibnamefont{Wang}},
  \bibinfo{author}{\bibfnamefont{K.}~\bibnamefont{Goodson}}, \bibnamefont{and}
  \bibinfo{author}{\bibfnamefont{H.}~\bibnamefont{Dai}},
  \bibinfo{journal}{Journal of Physical Chemistry B}
  \textbf{\bibinfo{volume}{110}}, \bibinfo{pages}{1502} (\bibinfo{year}{2006}),
  ISSN \bibinfo{issn}{15206106}, \eprint{0601139}.

\bibitem[{\citenamefont{Sullivan et~al.}(0)\citenamefont{Sullivan,
  Vallabhaneni, Kholmanov, Ruan, Murthy, and Shi}}]{Sullivan2017}
\bibinfo{author}{\bibfnamefont{S.}~\bibnamefont{Sullivan}},
  \bibinfo{author}{\bibfnamefont{A.}~\bibnamefont{Vallabhaneni}},
  \bibinfo{author}{\bibfnamefont{I.}~\bibnamefont{Kholmanov}},
  \bibinfo{author}{\bibfnamefont{X.}~\bibnamefont{Ruan}},
  \bibinfo{author}{\bibfnamefont{J.}~\bibnamefont{Murthy}}, \bibnamefont{and}
  \bibinfo{author}{\bibfnamefont{L.}~\bibnamefont{Shi}}, \bibinfo{journal}{Nano
  Letters} \textbf{\bibinfo{volume}{0}}, \bibinfo{pages}{null}
  (\bibinfo{year}{0}), \bibinfo{note}{pMID: 28218545},
  \eprint{http://dx.doi.org/10.1021/acs.nanolett.7b00110},
  \urlprefix\url{http://dx.doi.org/10.1021/acs.nanolett.7b00110}.

\bibitem[{\citenamefont{Maassen and Lundstrom}(2016)}]{Maassen2016}
\bibinfo{author}{\bibfnamefont{J.}~\bibnamefont{Maassen}} \bibnamefont{and}
  \bibinfo{author}{\bibfnamefont{M.}~\bibnamefont{Lundstrom}},
  \bibinfo{journal}{Journal of Applied Physics} \textbf{\bibinfo{volume}{119}},
  \bibinfo{pages}{095102} (\bibinfo{year}{2016}), ISSN
  \bibinfo{issn}{10897550}, \eprint{1508.03864},
  \urlprefix\url{http://dx.doi.org/10.1063/1.4942836}.

\bibitem[{\citenamefont{Vallabhaneni
  et~al.}(2016{\natexlab{a}})\citenamefont{Vallabhaneni, Singh, Bao, Murthy,
  and Ruan}}]{Ajit_Raman}
\bibinfo{author}{\bibfnamefont{A.~K.} \bibnamefont{Vallabhaneni}},
  \bibinfo{author}{\bibfnamefont{D.}~\bibnamefont{Singh}},
  \bibinfo{author}{\bibfnamefont{H.}~\bibnamefont{Bao}},
  \bibinfo{author}{\bibfnamefont{J.}~\bibnamefont{Murthy}}, \bibnamefont{and}
  \bibinfo{author}{\bibfnamefont{X.}~\bibnamefont{Ruan}},
  \bibinfo{journal}{Phys. Rev. B} \textbf{\bibinfo{volume}{93}},
  \bibinfo{pages}{125432} (\bibinfo{year}{2016}{\natexlab{a}}),
  \urlprefix\url{http://link.aps.org/doi/10.1103/PhysRevB.93.125432}.

\bibitem[{\citenamefont{Ni et~al.}(2012)\citenamefont{Ni, Aksamija, Murthy, and
  Ravaioli}}]{ni2012coupled}
\bibinfo{author}{\bibfnamefont{C.}~\bibnamefont{Ni}},
  \bibinfo{author}{\bibfnamefont{Z.}~\bibnamefont{Aksamija}},
  \bibinfo{author}{\bibfnamefont{J.~Y.} \bibnamefont{Murthy}},
  \bibnamefont{and} \bibinfo{author}{\bibfnamefont{U.}~\bibnamefont{Ravaioli}},
  \bibinfo{journal}{Journal of Computational Electronics}
  \textbf{\bibinfo{volume}{11}}, \bibinfo{pages}{93} (\bibinfo{year}{2012}).

\bibitem[{\citenamefont{Plimpton}(1995)}]{LAMMPS}
\bibinfo{author}{\bibfnamefont{S.}~\bibnamefont{Plimpton}},
  \bibinfo{journal}{Journal of Computational Physics}
  \textbf{\bibinfo{volume}{117}}, \bibinfo{pages}{1 } (\bibinfo{year}{1995}),
  ISSN \bibinfo{issn}{0021-9991},
  \urlprefix\url{http://www.sciencedirect.com/science/article/pii/S002199918571039X}.

\bibitem[{\citenamefont{Tersoff}(1989)}]{tersoff1989modeling}
\bibinfo{author}{\bibfnamefont{J.}~\bibnamefont{Tersoff}},
  \bibinfo{journal}{Physical Review B} \textbf{\bibinfo{volume}{39}},
  \bibinfo{pages}{5566} (\bibinfo{year}{1989}),
  \urlprefix\url{http://journals.aps.org/prb/abstract/10.1103/PhysRevB.39.5566}.

\bibitem[{\citenamefont{Tersoff}(1990)}]{tersoff1990erratum}
\bibinfo{author}{\bibfnamefont{J.}~\bibnamefont{Tersoff}},
  \bibinfo{journal}{Physical Review B} \textbf{\bibinfo{volume}{41}},
  \bibinfo{pages}{3248} (\bibinfo{year}{1990}),
  \urlprefix\url{http://journals.aps.org/prb/abstract/10.1103/PhysRevB.41.3248.2}.

\bibitem[{\citenamefont{Lindsay and Broido}(2010)}]{Lindsay_Tersoff}
\bibinfo{author}{\bibfnamefont{L.}~\bibnamefont{Lindsay}} \bibnamefont{and}
  \bibinfo{author}{\bibfnamefont{D.~A.} \bibnamefont{Broido}},
  \bibinfo{journal}{Physical Review B} \textbf{\bibinfo{volume}{81}},
  \bibinfo{pages}{205441} (\bibinfo{year}{2010}), ISSN
  \bibinfo{issn}{1098-0121},
  \urlprefix\url{http://link.aps.org/doi/10.1103/PhysRevB.81.205441}.

\bibitem[{\citenamefont{K{\i}nac{\i} et~al.}(2012)\citenamefont{K{\i}nac{\i},
  Haskins, Sevik, and {\c{C}}a{\u{g}}{\i}n}}]{kinaci2012thermal}
\bibinfo{author}{\bibfnamefont{A.}~\bibnamefont{K{\i}nac{\i}}},
  \bibinfo{author}{\bibfnamefont{J.~B.} \bibnamefont{Haskins}},
  \bibinfo{author}{\bibfnamefont{C.}~\bibnamefont{Sevik}}, \bibnamefont{and}
  \bibinfo{author}{\bibfnamefont{T.}~\bibnamefont{{\c{C}}a{\u{g}}{\i}n}},
  \bibinfo{journal}{Physical Review B} \textbf{\bibinfo{volume}{86}},
  \bibinfo{pages}{115410} (\bibinfo{year}{2012}),
  \urlprefix\url{http://journals.aps.org/prb/abstract/10.1103/PhysRevB.86.115410}.

\bibitem[{\citenamefont{Rapp{\'e} et~al.}(1992)\citenamefont{Rapp{\'e},
  Casewit, Colwell, Goddard~Iii, and Skiff}}]{rappe1992uff}
\bibinfo{author}{\bibfnamefont{A.~K.} \bibnamefont{Rapp{\'e}}},
  \bibinfo{author}{\bibfnamefont{C.~J.} \bibnamefont{Casewit}},
  \bibinfo{author}{\bibfnamefont{K.}~\bibnamefont{Colwell}},
  \bibinfo{author}{\bibfnamefont{W.}~\bibnamefont{Goddard~Iii}},
  \bibnamefont{and} \bibinfo{author}{\bibfnamefont{W.}~\bibnamefont{Skiff}},
  \bibinfo{journal}{Journal of the American chemical society}
  \textbf{\bibinfo{volume}{114}}, \bibinfo{pages}{10024}
  (\bibinfo{year}{1992}),
  \urlprefix\url{http://pubs.acs.org/doi/abs/10.1021/ja00051a040}.

\bibitem[{\citenamefont{Nicklow et~al.}(1972)\citenamefont{Nicklow,
  Wakabayashi, and Smith}}]{nicklow1972lattice}
\bibinfo{author}{\bibfnamefont{R.}~\bibnamefont{Nicklow}},
  \bibinfo{author}{\bibfnamefont{N.}~\bibnamefont{Wakabayashi}},
  \bibnamefont{and} \bibinfo{author}{\bibfnamefont{H.}~\bibnamefont{Smith}},
  \bibinfo{journal}{Physical Review B} \textbf{\bibinfo{volume}{5}},
  \bibinfo{pages}{4951} (\bibinfo{year}{1972}),
  \urlprefix\url{http://journals.aps.org/prb/abstract/10.1103/PhysRevB.5.4951}.

\bibitem[{\citenamefont{Lindsay et~al.}(2011)\citenamefont{Lindsay, Broido, and
  Mingo}}]{lindsay2011flexural}
\bibinfo{author}{\bibfnamefont{L.}~\bibnamefont{Lindsay}},
  \bibinfo{author}{\bibfnamefont{D.~A.} \bibnamefont{Broido}},
  \bibnamefont{and} \bibinfo{author}{\bibfnamefont{N.}~\bibnamefont{Mingo}},
  \bibinfo{journal}{Phys. Rev. B} \textbf{\bibinfo{volume}{83}},
  \bibinfo{pages}{235428} (\bibinfo{year}{2011}),
  \urlprefix\url{http://link.aps.org/doi/10.1103/PhysRevB.83.235428}.

\bibitem[{\citenamefont{Jeong et~al.}(2010)\citenamefont{Jeong, Kim, Luisier,
  Datta, and Lundstrom}}]{Jeong2010jap}
\bibinfo{author}{\bibfnamefont{C.}~\bibnamefont{Jeong}},
  \bibinfo{author}{\bibfnamefont{R.}~\bibnamefont{Kim}},
  \bibinfo{author}{\bibfnamefont{M.}~\bibnamefont{Luisier}},
  \bibinfo{author}{\bibfnamefont{S.}~\bibnamefont{Datta}}, \bibnamefont{and}
  \bibinfo{author}{\bibfnamefont{M.}~\bibnamefont{Lundstrom}},
  \bibinfo{journal}{Journal of Applied Physics} \textbf{\bibinfo{volume}{107}},
  \bibinfo{pages}{023707} (\bibinfo{year}{2010}), ISSN
  \bibinfo{issn}{00218979}, \eprint{0909.5222}.

\bibitem[{\citenamefont{Kaiser et~al.}(2016)\citenamefont{Kaiser, Feng,
  Maassen, Wang, Ruan, and Lundstrom}}]{Lundstrom2016}
\bibinfo{author}{\bibfnamefont{J.}~\bibnamefont{Kaiser}},
  \bibinfo{author}{\bibfnamefont{T.}~\bibnamefont{Feng}},
  \bibinfo{author}{\bibfnamefont{J.}~\bibnamefont{Maassen}},
  \bibinfo{author}{\bibfnamefont{X.}~\bibnamefont{Wang}},
  \bibinfo{author}{\bibfnamefont{X.}~\bibnamefont{Ruan}}, \bibnamefont{and}
  \bibinfo{author}{\bibfnamefont{M.}~\bibnamefont{Lundstrom}},
  \bibinfo{journal}{arXiv:1608.01188}  (\bibinfo{year}{2016}),
  \urlprefix\url{http://arxiv.org/abs/1608.01188}.

\bibitem[{\citenamefont{Zhou et~al.}(2007)\citenamefont{Zhou, Gweon, Fedorov,
  First, De~Heer, Lee, Guinea, Neto, and Lanzara}}]{Zhou2007Nm}
\bibinfo{author}{\bibfnamefont{S.}~\bibnamefont{Zhou}},
  \bibinfo{author}{\bibfnamefont{G.-H.} \bibnamefont{Gweon}},
  \bibinfo{author}{\bibfnamefont{A.}~\bibnamefont{Fedorov}},
  \bibinfo{author}{\bibfnamefont{P.}~\bibnamefont{First}},
  \bibinfo{author}{\bibfnamefont{W.}~\bibnamefont{De~Heer}},
  \bibinfo{author}{\bibfnamefont{D.-H.} \bibnamefont{Lee}},
  \bibinfo{author}{\bibfnamefont{F.}~\bibnamefont{Guinea}},
  \bibinfo{author}{\bibfnamefont{A.~C.} \bibnamefont{Neto}}, \bibnamefont{and}
  \bibinfo{author}{\bibfnamefont{A.}~\bibnamefont{Lanzara}},
  \bibinfo{journal}{Nature materials} \textbf{\bibinfo{volume}{6}},
  \bibinfo{pages}{770} (\bibinfo{year}{2007}).

\bibitem[{\citenamefont{Ong and Pop}(2011)}]{Ong2011prb}
\bibinfo{author}{\bibfnamefont{Z.-Y.} \bibnamefont{Ong}} \bibnamefont{and}
  \bibinfo{author}{\bibfnamefont{E.}~\bibnamefont{Pop}},
  \bibinfo{journal}{Physical Review B} \textbf{\bibinfo{volume}{84}},
  \bibinfo{pages}{075471} (\bibinfo{year}{2011}).

\bibitem[{\citenamefont{Qiu and Ruan}(2012)}]{Qiu2012apl2}
\bibinfo{author}{\bibfnamefont{B.}~\bibnamefont{Qiu}} \bibnamefont{and}
  \bibinfo{author}{\bibfnamefont{X.}~\bibnamefont{Ruan}},
  \bibinfo{journal}{Applied Physics Letters} \textbf{\bibinfo{volume}{100}},
  \bibinfo{pages}{193101} (\bibinfo{year}{2012}), ISSN
  \bibinfo{issn}{00036951},
  \urlprefix\url{http://scitation.aip.org/content/aip/journal/apl/100/19/10.1063/1.4712041}.

\bibitem[{\citenamefont{Fratini and Guinea}(2008)}]{Fratini2008prb}
\bibinfo{author}{\bibfnamefont{S.}~\bibnamefont{Fratini}} \bibnamefont{and}
  \bibinfo{author}{\bibfnamefont{F.}~\bibnamefont{Guinea}},
  \bibinfo{journal}{Physical Review B} \textbf{\bibinfo{volume}{77}},
  \bibinfo{pages}{195415} (\bibinfo{year}{2008}).

\bibitem[{\citenamefont{Chen et~al.}(2008)\citenamefont{Chen, Jang, Xiao,
  Ishigami, and Fuhrer}}]{Chen2008Nn}
\bibinfo{author}{\bibfnamefont{J.-H.} \bibnamefont{Chen}},
  \bibinfo{author}{\bibfnamefont{C.}~\bibnamefont{Jang}},
  \bibinfo{author}{\bibfnamefont{S.}~\bibnamefont{Xiao}},
  \bibinfo{author}{\bibfnamefont{M.}~\bibnamefont{Ishigami}}, \bibnamefont{and}
  \bibinfo{author}{\bibfnamefont{M.~S.} \bibnamefont{Fuhrer}},
  \bibinfo{journal}{Nature nanotechnology} \textbf{\bibinfo{volume}{3}},
  \bibinfo{pages}{206} (\bibinfo{year}{2008}).

\bibitem[{\citenamefont{Morozov et~al.}(2008)\citenamefont{Morozov, Novoselov,
  Katsnelson, Schedin, Elias, Jaszczak, and Geim}}]{Morozov2008prl}
\bibinfo{author}{\bibfnamefont{S.}~\bibnamefont{Morozov}},
  \bibinfo{author}{\bibfnamefont{K.}~\bibnamefont{Novoselov}},
  \bibinfo{author}{\bibfnamefont{M.}~\bibnamefont{Katsnelson}},
  \bibinfo{author}{\bibfnamefont{F.}~\bibnamefont{Schedin}},
  \bibinfo{author}{\bibfnamefont{D.}~\bibnamefont{Elias}},
  \bibinfo{author}{\bibfnamefont{J.}~\bibnamefont{Jaszczak}}, \bibnamefont{and}
  \bibinfo{author}{\bibfnamefont{A.}~\bibnamefont{Geim}},
  \bibinfo{journal}{Physical review letters} \textbf{\bibinfo{volume}{100}},
  \bibinfo{pages}{016602} (\bibinfo{year}{2008}).

\bibitem[{\citenamefont{Freitag et~al.}(2009)\citenamefont{Freitag, Steiner,
  Martin, Perebeinos, Chen, Tsang, and Avouris}}]{Freitag2009NL}
\bibinfo{author}{\bibfnamefont{M.}~\bibnamefont{Freitag}},
  \bibinfo{author}{\bibfnamefont{M.}~\bibnamefont{Steiner}},
  \bibinfo{author}{\bibfnamefont{Y.}~\bibnamefont{Martin}},
  \bibinfo{author}{\bibfnamefont{V.}~\bibnamefont{Perebeinos}},
  \bibinfo{author}{\bibfnamefont{Z.}~\bibnamefont{Chen}},
  \bibinfo{author}{\bibfnamefont{J.~C.} \bibnamefont{Tsang}}, \bibnamefont{and}
  \bibinfo{author}{\bibfnamefont{P.}~\bibnamefont{Avouris}},
  \bibinfo{journal}{Nano letters} \textbf{\bibinfo{volume}{9}},
  \bibinfo{pages}{1883} (\bibinfo{year}{2009}).

\bibitem[{\citenamefont{Yan et~al.}(2012)\citenamefont{Yan, Liu, Khan, and
  Balandin}}]{Yan2012NC}
\bibinfo{author}{\bibfnamefont{Z.}~\bibnamefont{Yan}},
  \bibinfo{author}{\bibfnamefont{G.}~\bibnamefont{Liu}},
  \bibinfo{author}{\bibfnamefont{J.~M.} \bibnamefont{Khan}}, \bibnamefont{and}
  \bibinfo{author}{\bibfnamefont{A.~A.} \bibnamefont{Balandin}},
  \bibinfo{journal}{Nature communications} \textbf{\bibinfo{volume}{3}},
  \bibinfo{pages}{827} (\bibinfo{year}{2012}).

\bibitem[{\citenamefont{Sadeghi et~al.}(2013)\citenamefont{Sadeghi, Jo, and
  Shi}}]{Sadeghi2013NL}
\bibinfo{author}{\bibfnamefont{M.~M.} \bibnamefont{Sadeghi}},
  \bibinfo{author}{\bibfnamefont{I.}~\bibnamefont{Jo}}, \bibnamefont{and}
  \bibinfo{author}{\bibfnamefont{L.}~\bibnamefont{Shi}},
  \bibinfo{journal}{Proceedings of the National Academy of Sciences of the
  United States of America} \textbf{\bibinfo{volume}{110}},
  \bibinfo{pages}{16321} (\bibinfo{year}{2013}), ISSN
  \bibinfo{issn}{1091-6490},
  \urlprefix\url{http://www.pnas.org/content/110/41/16321.full.pdf}.

\bibitem[{\citenamefont{Vallabhaneni
  et~al.}(2016{\natexlab{b}})\citenamefont{Vallabhaneni, Singh, Bao, Murthy,
  and Ruan}}]{Ajit2016prb}
\bibinfo{author}{\bibfnamefont{A.~K.} \bibnamefont{Vallabhaneni}},
  \bibinfo{author}{\bibfnamefont{D.}~\bibnamefont{Singh}},
  \bibinfo{author}{\bibfnamefont{H.}~\bibnamefont{Bao}},
  \bibinfo{author}{\bibfnamefont{J.}~\bibnamefont{Murthy}}, \bibnamefont{and}
  \bibinfo{author}{\bibfnamefont{X.}~\bibnamefont{Ruan}},
  \bibinfo{journal}{Phys. Rev. B} \textbf{\bibinfo{volume}{93}},
  \bibinfo{pages}{125432} (\bibinfo{year}{2016}{\natexlab{b}}),
  \urlprefix\url{http://link.aps.org/doi/10.1103/PhysRevB.93.125432}.

\bibitem[{\citenamefont{Majumdar and Reddy}(2004)}]{Majumdar2004apl}
\bibinfo{author}{\bibfnamefont{A.}~\bibnamefont{Majumdar}} \bibnamefont{and}
  \bibinfo{author}{\bibfnamefont{P.}~\bibnamefont{Reddy}},
  \bibinfo{journal}{Applied Physics Letters} \textbf{\bibinfo{volume}{84}}
  (\bibinfo{year}{2004}).

\bibitem[{\citenamefont{Wang et~al.}(2016)\citenamefont{Wang, Lu, Roy, and
  Ruan}}]{Wang2016jap}
\bibinfo{author}{\bibfnamefont{Y.}~\bibnamefont{Wang}},
  \bibinfo{author}{\bibfnamefont{Z.}~\bibnamefont{Lu}},
  \bibinfo{author}{\bibfnamefont{A.~K.} \bibnamefont{Roy}}, \bibnamefont{and}
  \bibinfo{author}{\bibfnamefont{X.}~\bibnamefont{Ruan}},
  \bibinfo{journal}{Journal of Applied Physics} \textbf{\bibinfo{volume}{119}},
  \bibinfo{eid}{065103} (\bibinfo{year}{2016}),
  \urlprefix\url{http://scitation.aip.org/content/aip/journal/jap/119/6/10.1063/1.4941347}.

\bibitem[{\citenamefont{Zhao et~al.}(2014)\citenamefont{Zhao, Hu, Song, Wang,
  Shi, Dai, and Qu}}]{Zhao2014EES}
\bibinfo{author}{\bibfnamefont{Y.}~\bibnamefont{Zhao}},
  \bibinfo{author}{\bibfnamefont{C.}~\bibnamefont{Hu}},
  \bibinfo{author}{\bibfnamefont{L.}~\bibnamefont{Song}},
  \bibinfo{author}{\bibfnamefont{L.}~\bibnamefont{Wang}},
  \bibinfo{author}{\bibfnamefont{G.}~\bibnamefont{Shi}},
  \bibinfo{author}{\bibfnamefont{L.}~\bibnamefont{Dai}}, \bibnamefont{and}
  \bibinfo{author}{\bibfnamefont{L.}~\bibnamefont{Qu}},
  \bibinfo{journal}{Energy \& Environmental Science}
  \textbf{\bibinfo{volume}{7}}, \bibinfo{pages}{1913} (\bibinfo{year}{2014}).

\bibitem[{\citenamefont{Zhou et~al.}(2013)\citenamefont{Zhou, Lin, Huang,
  Zhong, Wang, Tang, Bi, Wan, and Lin}}]{Zhou2013AFM}
\bibinfo{author}{\bibfnamefont{M.}~\bibnamefont{Zhou}},
  \bibinfo{author}{\bibfnamefont{T.}~\bibnamefont{Lin}},
  \bibinfo{author}{\bibfnamefont{F.}~\bibnamefont{Huang}},
  \bibinfo{author}{\bibfnamefont{Y.}~\bibnamefont{Zhong}},
  \bibinfo{author}{\bibfnamefont{Z.}~\bibnamefont{Wang}},
  \bibinfo{author}{\bibfnamefont{Y.}~\bibnamefont{Tang}},
  \bibinfo{author}{\bibfnamefont{H.}~\bibnamefont{Bi}},
  \bibinfo{author}{\bibfnamefont{D.}~\bibnamefont{Wan}}, \bibnamefont{and}
  \bibinfo{author}{\bibfnamefont{J.}~\bibnamefont{Lin}},
  \bibinfo{journal}{Advanced Functional Materials}
  \textbf{\bibinfo{volume}{23}}, \bibinfo{pages}{2263} (\bibinfo{year}{2013}).

\bibitem[{\citenamefont{Han et~al.}(2014)\citenamefont{Han, Wu, Li, Zhang, and
  Feng}}]{Han2014AM}
\bibinfo{author}{\bibfnamefont{S.}~\bibnamefont{Han}},
  \bibinfo{author}{\bibfnamefont{D.}~\bibnamefont{Wu}},
  \bibinfo{author}{\bibfnamefont{S.}~\bibnamefont{Li}},
  \bibinfo{author}{\bibfnamefont{F.}~\bibnamefont{Zhang}}, \bibnamefont{and}
  \bibinfo{author}{\bibfnamefont{X.}~\bibnamefont{Feng}},
  \bibinfo{journal}{Advanced Materials} \textbf{\bibinfo{volume}{26}},
  \bibinfo{pages}{849} (\bibinfo{year}{2014}).

\bibitem[{\citenamefont{Huang et~al.}(2014)\citenamefont{Huang, Tan, Yin, and
  Zhang}}]{Huang2014AM}
\bibinfo{author}{\bibfnamefont{X.}~\bibnamefont{Huang}},
  \bibinfo{author}{\bibfnamefont{C.}~\bibnamefont{Tan}},
  \bibinfo{author}{\bibfnamefont{Z.}~\bibnamefont{Yin}}, \bibnamefont{and}
  \bibinfo{author}{\bibfnamefont{H.}~\bibnamefont{Zhang}},
  \bibinfo{journal}{Advanced Materials} \textbf{\bibinfo{volume}{26}},
  \bibinfo{pages}{2185} (\bibinfo{year}{2014}).

\bibitem[{\citenamefont{Thiyagarajan et~al.}(2014)\citenamefont{Thiyagarajan,
  Oh, Yoon, and Jang}}]{Thiyagarajan2014APL}
\bibinfo{author}{\bibfnamefont{P.}~\bibnamefont{Thiyagarajan}},
  \bibinfo{author}{\bibfnamefont{M.-W.} \bibnamefont{Oh}},
  \bibinfo{author}{\bibfnamefont{J.-C.} \bibnamefont{Yoon}}, \bibnamefont{and}
  \bibinfo{author}{\bibfnamefont{J.-H.} \bibnamefont{Jang}},
  \bibinfo{journal}{Applied Physics Letters} \textbf{\bibinfo{volume}{105}},
  \bibinfo{pages}{033905} (\bibinfo{year}{2014}).

\bibitem[{\citenamefont{Qiu and Ruan}(2011)}]{Qiu2011arXiv}
\bibinfo{author}{\bibfnamefont{B.}~\bibnamefont{Qiu}} \bibnamefont{and}
  \bibinfo{author}{\bibfnamefont{X.}~\bibnamefont{Ruan}},
  \bibinfo{journal}{arXiv preprint arXiv:1111.4613}  (\bibinfo{year}{2011}),
  \eprint{arXiv:1111.4613v1}, \urlprefix\url{http://arxiv.org/abs/1111.4613v1}.

\bibitem[{\citenamefont{Feng et~al.}(2015)\citenamefont{Feng, Qiu, and
  Ruan}}]{Feng2015jap}
\bibinfo{author}{\bibfnamefont{T.}~\bibnamefont{Feng}},
  \bibinfo{author}{\bibfnamefont{B.}~\bibnamefont{Qiu}}, \bibnamefont{and}
  \bibinfo{author}{\bibfnamefont{X.}~\bibnamefont{Ruan}},
  \bibinfo{journal}{Journal of Applied Physics} \textbf{\bibinfo{volume}{117}},
  \bibinfo{pages}{195102} (\bibinfo{year}{2015}), ISSN
  \bibinfo{issn}{0021-8979},
  \urlprefix\url{http://scitation.aip.org/content/aip/journal/jap/117/19/10.1063/1.4921108}.

\bibitem[{\citenamefont{de~Koker}(2009)}]{Koker2009prl}
\bibinfo{author}{\bibfnamefont{N.}~\bibnamefont{de~Koker}},
  \bibinfo{journal}{Physical Review Letters} \textbf{\bibinfo{volume}{103}},
  \bibinfo{pages}{125902} (\bibinfo{year}{2009}), ISSN
  \bibinfo{issn}{0031-9007},
  \urlprefix\url{http://link.aps.org/doi/10.1103/PhysRevLett.103.125902}.

\bibitem[{\citenamefont{Thomas et~al.}(2010)\citenamefont{Thomas, Turney,
  Iutzi, Amon, and McGaughey}}]{Thomas2010prb}
\bibinfo{author}{\bibfnamefont{J.~A.} \bibnamefont{Thomas}},
  \bibinfo{author}{\bibfnamefont{J.~E.} \bibnamefont{Turney}},
  \bibinfo{author}{\bibfnamefont{R.~M.} \bibnamefont{Iutzi}},
  \bibinfo{author}{\bibfnamefont{C.~H.} \bibnamefont{Amon}}, \bibnamefont{and}
  \bibinfo{author}{\bibfnamefont{A.~J.~H.} \bibnamefont{McGaughey}},
  \bibinfo{journal}{Physical Review B} \textbf{\bibinfo{volume}{81}},
  \bibinfo{pages}{081411} (\bibinfo{year}{2010}), ISSN
  \bibinfo{issn}{1098-0121},
  \urlprefix\url{http://link.aps.org/doi/10.1103/PhysRevB.81.081411}.

\end{thebibliography}


\end{document}